\documentclass[twocolumn]{aa} 
\usepackage{longtable}
\usepackage{natbib}
\bibpunct{(}{)}{;}{a}{}{,} 

\usepackage{graphicx}
\usepackage{txfonts}
\usepackage{verbatim}
\begin{document}
   \title{Circumstellar disks around Herbig Be stars}
   \author{T. Alonso-Albi
          \inst{1}
          \and
          A. Fuente
          \inst{1}
	\and
	R. Bachiller
	\inst{1}
	\and
	R. Neri
	\inst{2}
	\and
	P. Planesas
	\inst{1, 3}
        \and
        L. Testi
        \inst{4, 6}
          \and
          O. Bern{\'e}
          \inst{5}
          \and
          C. Joblin
          \inst{5}
          }

   \offprints{t.alonso@oan.es}

   \institute{Observatorio Astron\'omico Nacional, Apdo. 112,
              E-28803 Alcal\'a de Henares (Madrid), Spain 
        \and 
             Institut de Radio Astronomie Milimetrique, 300 Rue de la Piscine, Domaine Universitaire de Grenoble, F-38406 St. Martin d'H\`{e}res, France
         \and
         Atacama Large Millimeter/Submillimeter Array, Joint ALMA Office, Santiago, Chile
        \and
             INAF-Osservatorio Astrofisico de Arcetri, Largo Enrico Fermi 5, I-50125 Firenze, Italy
         \and
         Centre d'\'Etude Spatiale des Rayonnements, CNRS et Universit\'e Paul Sabatier Toulouse 3, Observatoire Midi-Pyr\'en\'ees, 9 Av. du Colonel Roche, 31028 Toulouse Cedex 04, France
	\and
         European Southern Observatory, Karl Schwarzschild str. 2, D-85748, Garching, Germany
}
   \date{Received ; accepted }

 
  \abstract
  {}
  {
Our goal is to investigate 
the properties of the circumstellar disks around intermediate mass stars to determine
their occurrence, lifetime and evolution. 
}
{We completed a search for circumstellar disks around Herbig Be stars using the NRAO 
Very Large Array (VLA) and the IRAM Plateau de Bure (PdB) interferometers. Thus far, we have observed 
6 objects with 4 successful detections.
The results towards 3 of these stars (R~Mon, MWC~1080, MWC~137) were presented elsewhere. 
We present our new VLA and PdBI data for the three objects MWC~297, Z~CMa, and LKH$\alpha$~215.
We constructed the SED from near-IR to centimeter wavelengths by adding our millimeter and 
centimeter data to the available data at other wavelengths, mainly Spitzer images. The entire SED
was fitted using a disk+envelope model. In addition, we compiled all the disk millimeter observations in the literature and completed a statistical analysis of all the data.
}
{
We show that the disk mass is usually only a small percentage (less than 10$\%$) of the mass of the entire envelope in HBe stars. For the disks, there are large source-to-source variations. Two disks in our sample, R~Mon and Z~CMa, have similar sizes and masses to those found in T Tauri and Herbig Ae stars. The disks around MWC~1080 and MWC~297 are, however, smaller (r$_{out}$$<$100~AU). We did not detect the disks towards MWC~137 and LkH$\alpha$~215 at millimeter wavelengths, which limits the mass and the size of the possible circumstellar disks.
}
{
A comparison between our data and previous results for T Tauri and Herbig Ae stars indicates that although massive disks ($\sim$0.1~M$_\odot$) are found in young objects ($\sim$10$^4$ yr), 
the masses of the disks around Herbig Be stars are usually 5-10 times lower than those around lower mass stars. We propose that disk photoevaporation is responsible for this behavior. In Herbig Be stars, the UV radiation disperses the gas in the outer disk on a timescale of a few 10$^5$ yr. Once the outer part of the disk has vanished, the entire gaseous disk is photoevaporated on a very short timescale ($\sim$10$^5$ yr) and only a small, dusty disk consisting of large grains remains.
}

   \keywords{stars: individual: MWC~137, LkH$\alpha$~215, R~Mon, Z~CMa, MWC~297, MWC~1080--stars: formation--stars:pre-main-sequence -- circumstellar matter -- planetary systems: formation -- planetary systems: protoplanetary disks}

   \maketitle
%

\section{Introduction}

Herbig Ae/Be (HAeBe) stars are intermediate-mass (M$\sim$2--8~M$_\odot$) pre-main-sequence objects. Since these objects share many characteristics with high-mass stars (clustering, PDRs), but are far closer to us and less embedded, the detection of circumstellar disks around these stars is crucial to the understanding of the massive-star-formation process. Herbig Ae (HAe) stars are the precursors of Vega-type systems: determining the frequency and timescales of the disks around HAe stars is therefore also important for planet formation studies.

Substantial theoretical and observational efforts have been carried out recently for the understanding of the disk occurrence and evolution in HAeBe stars \citep{mee01, vin02, mil01, ack05}. On the basis of the different mid-infrared excesses, \cite{mee01} proposed that disks around Herbig Ae/Be stars can be classified into two Groups that they interpreted in terms of different geometries: Group I sources with strong mid-infrared (20--100 $\mu$m) flux excesses and Group II sources with modest infrared excesses. Group I sources have Chiang \& Goldreich-like flared disks and Group II sources host flatter, self-shadowed disks. \cite{lei04} confirmed this hypothesis on the basis of 10~$\mu$m interferometry data. While most HAe stars belong to Group I and are surrounded by disks similar to those in T Tauri (TT) stars, Herbig Be (HBe) stars are mostly found in Group II and seem to be surrounded by disks with a flatter geometry \citep{ack05}.

Although the existence of different types of disks is well accepted, whether this difference is due to the different characteristics of the central star (hotter stars produce flatter disks), or to different characteristics of the circumstellar matter (different grain opacities), or to an evolutionary link between flared, self-shadowed, and debris disks is not established. If the evolutionary scenario stands, the different geometry between disks in HAe and HBe stars is the consequence of a more rapid grain growth and a shorter dissipation timescale in the latter. The grain growth causes the optical depth of the disk to decrease and allows the UV radiation to penetrate deep into the circumstellar disk and photoevaporate the disk external layers \citep{dul04}. Disks around HBe would flatten and lose a large fraction of their mass before the pre-main-sequence phase ($<$0.1~Myr).

One problem in discerning between the different scenarios is that most studies are based on the optical-NIR and mid-IR observations. Optical-NIR and mid-IR observations probe the region within a few AU around the star. In addition, since the disk is optically thick at these wavelengths, they only provide information about the disk surface and cannot be used to derive the disk mass. An obvious step towards the formation of planetesimals in disks is coagulation and growth of the submicron sized grains accreted from the proto-stellar envelope \citep{dom07}. Although the end results are plainly visible in our own planetary system, the details and mechanisms of dust evolution are not understood. The main problem is that the cooler, large grains do not emit significantly at infrared wavelengths. Only by observing the submm-mm range of the spectrum are we able to determine the sizes and properties of these large grains. 
Millimeter observations must be used to trace this more evolved grain component. Thus, imaging disks at mm wavelengths is required to trace the outer part of the disk and determine its mass and grain properties.

We have carried out a search for circumstellar disks around HBe stars using the NRAO 
Very Large Array (VLA) and the IRAM Plateau de Bure (PdB) interferometers to investigate 
the properties of the circumstellar disks around intermediate-mass stars and to determine eventually
their occurence, lifetime, and evolution. Some results have already been published elsewhere \citep{fue03,fue06,alo07,alo08}. 
In this Paper, we present new observational data (LkH$\alpha$~215, Z~CMa and MWC~297) that complete our 
survey and we provide a comprehensive analysis of all the data obtained thus far.

\section{Observations}

\subsection{Selected sample}

The list of observed sources is shown in Table \ref{sample}. Our sample was built according to the following 
criteria: Firstly, we included most well-studied Herbig stars with spectral type close to B0 in the northern hemisphere (R~Mon, MWC~1080, MWC~137, MWC~297). In a pioneering work, \cite{nat00} compiled the existing 1.3mm/2.7mm interferometric disk detections in HAeBe stars. By this time, only one disk had been detected in stars with spectral type earlier than B7--8. They proposed that this lack of detections could be due to a shorter dissipation timescale for the circumstellar disk in stars of spectral type earlier than B7. If the disk mass decreases with stellar spectral type, one expects to measure the minimum disk masses towards the most massive B0 stars. Secondly, we included some objects with spectral types between B7 and B8, where the transition between the TT-like and massive star disks, if it exists, is expected to occur. Additional criteria were the closeness to the Sun and previous detections with single-dish telescopes. 

\subsection{Millimeter and centimeter observations}
Thus far, we have observed 6 HBe stars with 4 successful detections.
The VLA and PdBI observations towards R Mon, MWC~1080, and MWC~137 were already reported 
by \cite{fue03,fue06}. In this Paper, we present new VLA observations towards MWC~297 and Z~CMa, 
and new PdBI observations towards LkH$\alpha$~215, Z~CMa and MWC~297. A summary of these observations, wavelength, beam, and date, is shown in Table \ref{observations}.

The data were calibrated, mapped, and analyzed in the GILDAS software package. Continuum images at mm wavelengths
have been produced  by averaging the channels free of line emission after careful visual inspection. 
Natural weightings were applied to the measured visibilities producing 
the beams shown in Table \ref{observations}. Since the disks are expected to be unresolved by our interferometric
observations, we used the peak flux to construct the SED. The measured peak fluxes are
shown in Table 3. The rms of the images are shown in Tables A.1, A.3, A.5, A.7, A.9 and A.11.

\begin{table}
\caption{Sample of observed Herbig Be stars}
\begin{tabular}
{@{\extracolsep{\fill}}llllll} 
\hline\hline\noalign{\smallskip}
Name & R.A. & Dec. & Dist. & Mass & Sp. \\ 
     & (J2000) & (J2000) & (pc) & (M$_\odot$) & type \\ 
\hline\noalign{\smallskip}
MWC~137         & 06:18:45.50 & 15:16:52.4  & 1300 & 14 & B0 \\ 
R~Mon           & 06:39:09.95 & 08:44:10.7  & 800 & 8 & B0 \\ 
MWC~1080        & 23:17:25.57 & 60:50:43.3  & 1000 & 10 & B0 \\ 
MWC~297         & 18:27:39.53 & -03:49:52.0 & 250 & 9 & B1.5 \\ 
LKH$\alpha$~215 & 06:32:41.79 & 10:09:33.6  & 800 & 7 & B7.5 \\ 
Z~CMa           & 07:03:43.16 & -11:33:06.2 & 930 & 12 & B8 \\ 
\noalign{\smallskip}\hline\hline
\label{sample}
\end{tabular}
\end{table}

\subsection{Mid-IR observations}

Four of the 6 sources in our sample have been observed at 24 $\mu$m either with MIPS or IRS onboard the Spitzer Space Telescope, namely MWC~1080, MWC~297, R~Mon, and LkH$\alpha$~215. We retrieved the post-basic-calibrated data (pbcd) for imagery and basic-calibrated data for spectroscopy (bcd). The pbcd files were used without additional processing. The bcd spectral mapping files were processed with the CUBISM software \citep{smi07}.

\section{Results}
\subsection{Interferometric data}
The results of our observations are shown in Table \ref{results}. We detected emission
at 1.3mm and/or 2.7mm towards 5 out of the 6 stars studied. Emission at centimeter (cm) wavelengths using the VLA array was detected towards Z~CMa, MWC~297, MWC~137, and MWC~1080. In these massive stars, a significant fraction of the 1.3mm flux can be attributed to free-free emission instead of dust, thermal emission arising in the circumstellar disk. To estimate this contribution, we fitted the free-free spectral index at cm wavelengths and extrapolated the emission to millimeter (mm) wavelengths. In the case of MWC~137,
all the flux seems to be due to free-free emission without any significant excess at 1.3mm and 2.7mm. In the other cases, there is some flux excess at mm wavelengths that can be interpreted as arising in a circumstellar disk. We show in Table \ref{results} the spectral index between 1.3mm and 2.7mm interferometric observations. Two values are presented: the first represents the total observed fluxes and the second with the free-free emission substracted, i.e., the spectral index of the thermal, dust emission. 
The 1.3mm/2.7mm spectral index is quite low ($<$2.2) for all sources. As discussed in the following Sections, this low spectral index constitutes observational evidence of either optically thick dust emission or the presence of large grains in the disk midplane (see e.g. Alonso-Albi et al. 2008). As we discuss in Sect. 5.7, the second possibility is the most plausible in the circumstellar disks around R~Mon and Z~CMa.

\label{inteferometricData}
In Table \ref{results}, we also show the ratio of our interferometric fluxes to previous single-dish data. In most cases, the flux detected by interferometric techniques is far lower ($<$10\%) than that detected by single-dish observations. This implies that the envelope emission dominates the single-dish flux at FIR and mm wavelengths in most HBe stars. For this reason, interferometric measurements are essential to estimating the true disk emission. Since interferometric observations are unavailable at all wavelengths, a disk+envelope model is required to fit the observed SED properly.

\begin{table}
\caption{Observations}
\begin{tabular}
{@{\extracolsep{\fill}}lllllllll}
\hline\hline\noalign{\smallskip}
Name & $\lambda$ (mm) & Beam size & Date & Telescope \\
\hline\noalign{\smallskip}
MWC~297 & 1.3 & 1.1$''$ $\times$ 0.4$''$ & Feb 2006 & PdBI \\
        & 2.6 & 1.4$''$ $\times$ 0.9$''$ & Feb 2006 & \\
LKH$\alpha$~215 & 1.3 & 3.3$''$ $\times$ 1.8$''$ & April 2002 & PdBI \\
                & 2.6 & 6.8$''$ $\times$ 3.7$''$ & April 2002 & \\
Z~CMa           & 1.3 & 2.3$''$ $\times$ 0.9$''$ & Jan/March 2006 & PdBI \\
                & 2.6 & 4.1$''$ $\times$ 2.0$''$ & Jan/March 2006 &  \\ \hline\noalign{\smallskip}
MWC 297 &  7  & 1.9$''$ $\times$ 1.6$''$ & Dec 2005 & VLA\\
        &  13 & 3.9$''$ $\times$ 3.1$''$ & Dec 2005 & \\
Z CMa   &  7  & 1.4$''$ $\times$ 1.4$''$ & Oct/Dec 2005 & VLA \\
        &  13 & 3.5$''$ $\times$ 2.7$''$ & Oct/Dec 2005 & \\
        &  36 & 9.2$''$ $\times$ 6.6$''$ & Oct/Dec 2005 &  \\
\noalign{\smallskip}\hline\hline
\end{tabular}
\label{observations}
\end{table}

\begin{table*}
\caption{Summary of observational results}
\begin{tabular}{llllllll}
\hline\hline\noalign{\smallskip}
Name & 3.6 cm & 1.3 cm & 7 mm & 2.7 mm & 1.3 mm & Spec. index$^1$ & F$_{PdBI}$ / F$^2_{JCMT}$ \\
 & (mJy/beam) & (mJy/beam) & (mJy/beam) & (mJy/beam) & (mJy/beam) & (1.3-2.6 mm) & (1.3 mm) \\
\hline\noalign{\smallskip}
MWC 137 &  & 0.9 & 1.3 & 4.1 & 7.1 & 0.79/- & 0.075 \\
R Mon &  & 0.9 & 1.3 & 4.1 & 11.8 & 1.53/2.14 & 0.15 \\
MWC 1080 &  &  &  & $<$1.7 & 3.1 & $>$0.87/$>$1.02 & 0.013 \\
MWC 297 & & 23 & 29 & 149 & 175 & 1.01$^3$/1.31$^3$ & 0.39 \\  
LkH$\alpha$~215 &  &  &  & $<$0.6 & $<$1.5 & -/- & $<$0.026 \\
Z CMa & 2.2 & 2.1 & 2.1 & 8.5 & 26 & 1.61/1.88 & 0.037 \\
\noalign{\smallskip}\hline\hline
\end{tabular}

\noindent
$^1$ The 1.3mm--2.7mm spectral index before ({\it left}) and after ({\it right}) subtracting the
free-free emission.\\
\noindent
$^2$ See references in Appendix A.\\
\noindent
$^3$ Beam dependent, see Section \ref{discuss297}.
\label{results}
\end{table*}

\subsection{Observed SED}
We reconstructed the entire SED towards our sample by completing our VLA, PdBI, and Spitzer photometry with the data available in the literature. At optical and NIR wavelengths, observations are heavily affected by extinction. To correct for this reddening, we substracted the standard B-V color corresponding to the spectral type of each star, as given by \cite{joh66}, to the observed B-V color. This color excess \textit{E} was used to calculate the visual extinction \textit{A$_v$ = R$_v$ E} assuming a ratio of total-to-selective absorption of $R_v$ = 3.1 \citep{car89}. With these two parameters, we estimate the extinction corresponding to a particular observation using the parameterized extinction law of \cite{car89}, which can be applied to wavelengths between 0.12 and 3.5 $\mu$m. The complete photometric set of values with the observed and dereddened fluxes for each source is presented in Appendix A.

Most HAeBe stars are well-known variable stars, a problem that is relevant to our case because we used photometry recovered from the literature in the past decades. The variability affects the B-V color and produces some additional uncertainty in the extinction estimates. 

\begin{table*}
\caption{Disk modeling results$^1$
} 
\begin{tabular*}{1.0\textwidth}
{@{\extracolsep{\fill}}lllllllll} 
\hline\hline\noalign{\smallskip}
Name & Dust & Inner radius & Rim T & Rim height & Density & Inclination & Outer radius & 
Dispersion \\
 & Mass (M$_\odot$) & (AU) & (K) & (AU) & slope & (degrees) & (AU) & \\ 
\hline\noalign{\smallskip}
MWC~137  & 1.0~x10$^{-5}$ $^{2x10^{-4}}_{5x10^{-6}}$ & 8.4 & & & -1 $^{-}_{> -1.2}$ & 80 $^{90}_{70}$ & 18 $^{25}_{-}$ & 0.20 \\ 
R~Mon    & 1.4~x10$^{-4}$ $^{2x10^{-4}}_{9x10^{-5}}$ & 18.2 & & & -0.45 $^{0.0}_{-1.0}$ & 60 $^{70}_{40}$ & 150 $^{200}_{100}$ & 0.09 \\ 
MWC~1080 & 5.0~x10$^{-5}$ $^{1x10^{-4}}_{2.5x10^{-5}}$ & 5.6  & 1500 & 1.0 & -0.55 $^{-0.2}_{-0.8}$ & 83 $^{90}_{50}$ & 77 $^{100}_{60}$ & 0.08 \\ 
MWC~297  & 5.3~x10$^{-4}$ $^{8.0x10^{-4}}_{4.5x10^{-4}}$ & 5.0 & 1800 & 1.2 & -0.77 $^{-0.6}_{-0.9}$ & 5 $^{10}_{0}$ & 28.5 $^{30}_{25}$ & 0.19 \\ 
         & 4.0~x10$^{-4}$ $^{6.0x10^{-4}}_{3x10^{-4}}$ & 200 $^{100}_{300}$ &      &     &    &  & 300 $^{200}_{400}$ & \\ 
LKH$\alpha$~215 & 6.0~x10$^{-8}$ $^{9x10^{-8}}_{4x10^{-8}}$ & 3.5 & & & -1.4 $^{-1.2}_{-1.8}$ & 30 $^{50}_{0}$ & 9.2 $^{20}_{-}$ & 0.1 \\ 
Z~CMa           & 7.0~x10$^{-4}$ $^{1.4x10^{-3}}_{3.5x10^{-4}}$ & 7.5 & 1500 & 1.9 & -0.7 $^{-}_{>-1.0}$ & 30 $^{40}_{20}$ & 180 $^{250}_{140}$ & 0.15 \\ 
\noalign{\smallskip}\hline\hline
\end{tabular*}
\label{disks}

\noindent
$^1$ In the case of fitted parameters, we provide in small figures the range of allowed values. These values correspond to numerical errors. See Sect.~5.7 for further discussion of the uncertainties in the disk parameters.

\end{table*}

\begin{table}
\caption{Grain properties in the disks$^1$.} 
\begin{tabular}
{@{\extracolsep{\fill}}lll}
\hline\hline\noalign{\smallskip}
Name & Midplane & Surface \\
 & grains & grains  \\
\hline\noalign{\smallskip}
MWC~137  & SM  a$_{max}$=1~cm    & SM a$_{max}$=1~$\mu$m  \\
R~Mon    & SM  a$_{max}$=1~cm    & 86\% graphite $^{100\%}_{50\%}$  a$_{max}$=1~$\mu$m       \\
MWC~1080 & SM  a$_{max}$=1~cm    & SM a$_{max}$=100~$\mu$m  \\
MWC~297  & SM  a$_{max}$=1~cm    & SM a$_{max}$=1~$\mu$m  \\
LKHa~215 & SM  a$_{max}$=1~cm    & SM a$_{max}$=100~$\mu$m  \\
Z~CMa    & SM  a$_{max}$=1~cm   & 60\% graphite$^{100\%}_{50\%}$ a$_{max}$=1~$\mu$m  \\
\noalign{\smallskip}\hline\hline
\end{tabular}

\noindent
$^1$ The index of the power-law grain-size distribution is fixed (p=3.5). As a first step, we assumed the standard mixture (SM: 86\% silicates/14\% graphite) in the midplane and surface layers in all the sources. In the cases of R~Mon and Z~CMa, we needed to change the dust composition in the surface to fit the SED. The range of allowed values are indicated (small figures) in the Table. See Sect.~5.7 for a detailed discussion of the uncertainties.
\label{grains}
\end{table}

\begin{table*}
\caption{Envelope modeling results.} 
\begin{tabular*}{1.0\textwidth}
{@{\extracolsep{\fill}}llllllllll}
\hline\hline\noalign{\smallskip}
Name & Dust mass & Inner radius & Outer radius & Density & Inner temperature & Geometry & Inclination \\
     & (M$_\odot$) & (AU) & (AU) & slope & (K) & & (degrees) \\
\hline\noalign{\smallskip}
MWC~137 & 0.005 & 5000.0 & 11700.0 & -1.0 & 61 & Sphere &  \\
R~Mon & 0.008 & 700.0 & 12000.0 & -1.4 & 62 & Toroid$^1$ & 70.0 \\
MWC~1080 & 0.025 & 6000.0 & 12000.0 & -1.0 & 44 & Toroid & 80.0 \\
Z~CMa & 0.05 & 2000.0 & 5000.0 & -0.6 & 42 & Toroid$^1$ & 30.0  \\
LKH$\alpha$~215 & 0.0015 & 3500.0 & 7200.0 & 0.0 & 57 & Sphere &  \\
\noalign{\smallskip}\hline\hline
\end{tabular*}
\noindent
$^1$ The dust temperature in the envelope is assumed to be described by a power law, $T(r) = T_{in} (r/r_{in})^{-0.5}$. See Sections 4 and 5 for a more detailed discussion.
\label{envelopes}
\end{table*}

\section{Model}
\label{model}

As mentioned in Sect. \ref{inteferometricData}, a two-component model is required to fit the observed SEDs. The disks were fitted using a passive, irradiated circumstellar disk model described by \cite{alo08} allowing us to consider different grain populations for the midplane and surface layers. Grain populations were characterized by a silicate/graphite mixture, the maximum grain size ($a_{max}$), and the slope of the grain size distribution ($p$). The value of $p$ was fixed to 3.5 and a standard grain mixture (SM: 86\% silicate/14\% graphite) were assumed for the midplane and surface layers. In two cases, R~Mon and Z~CMa, we were unable to obtain a reliable fitting with these assumptions and we needed to vary the grain mixture in the surface layer. These cases are discussed in detail in Sect.~5.7.

In the fitting process, we consider that the emission at wavelengths shorter than 10~$\mu$m and the fluxes measured by interferometric techniques (PdBI,VLA) arise only in the star/disk system. This is based on the folowing scientific arguments: Firstly, the dust emitting at $\leq$10 $\mu$m has a temperature of $>$300K. Assuming that the star is the
only heating source in the region, the dust should be located at radii $<$ 100--300 AU, which is the typical size of a circumstellar disk. Secondly, the spatial resolution of our interferometric mm and cm observations is usually below 1000~AU. It is, again, reasonable to consider that most of the emission detected with the interferometers arise in the disk, although a contribution from the envelope may also exist in some cases.
The emission observed at mid-IR and FIR wavelengths (Spitzer, SCUBA) are, however, assumed to be the sum of the disk and the envelope components. This is consistent with the MIPS images at 24 $\mu$m, which show nebulous emission around MWC~1080, MWC~297, R~Mon, and LkH$\alpha$~215. 

The envelope was modeled using a ray-tracing code. The dust mass, density law, and geometry were determined by the fitting procedure. Our aim was to fit the observed emission and the visual extinction towards the star. Since the visual extinction was determined by the amount of mass along the line-of-sight, it constrains the geometry of the envelope. As discussed in Sect. \ref{individualSources}, the morphology of the continuum submillimeter and millimeter maps were also taken into account in constraining the envelope geometry. We modeled each envelope by considering two possible geometries, a sphere and a toroid. The dust temperature was assumed to be described by a power law, $T(r) = T_o (r/r_o)^{-0.5}$, where the values of $T_o$ and $r_o$ depends on whether the envelope was shadowed or not by the inner circumstellar disk. In the non-shadowed case, typical of a sphere geometry, the effective temperature and radius of the star were used as values for T$_o$ and r$_o$. In the shadowed case, typical of toroidal envelopes, we extrapolated the dust temperature in the disk midplane.
The dust opacities in the envelope were assumed to be those reported by \cite{oss94} for protostellar cores, using the standard MRN dust model with icy mantles.
The resulting emission was convolved by the beam of the observations before being compared with the observed SED.

The goodness of fit was estimated using a dispersion parameter ($\chi$) defined to be $\chi~=~\sqrt{\sum{((F_{observed}-F_{modeled})/F_{observed})^2}/n}$, where \textit{F} is the flux and the sum is extended for all detections in the SED dominated by the disk emission from 2~$\mu$m and beyond, and \textit{n} the number of detections. In Table \ref{disks}, we show the best-fit (minimum value of $\chi$) solution for each source.

\begin{figure*}
 \includegraphics[viewport=0 -30 500 700, width=0.4\textwidth, height=0.5\textwidth, angle=-90]{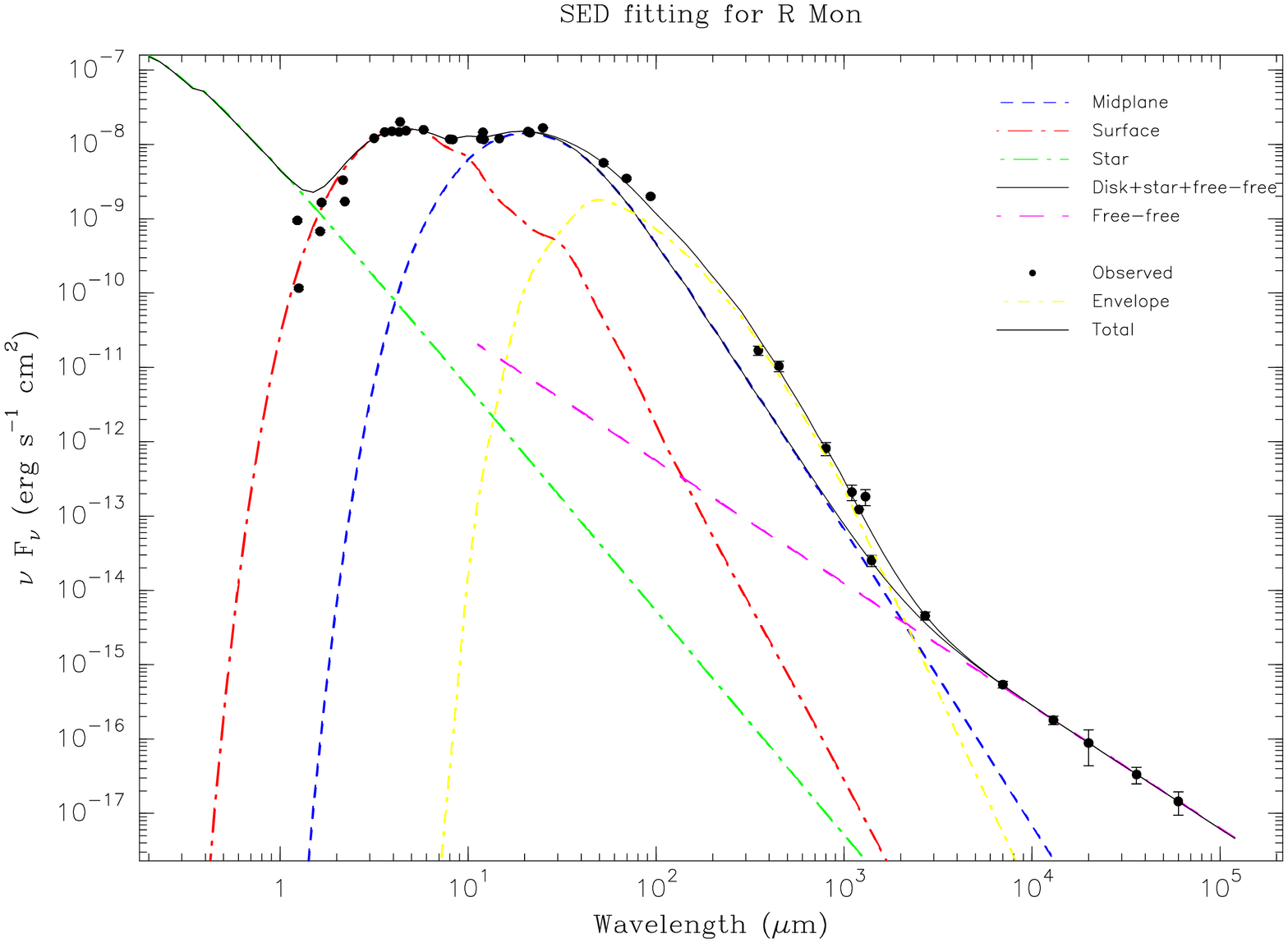}
 \includegraphics[viewport=-40 -100 620 450, width=0.4\textwidth, height=0.5\textwidth, angle=-90]{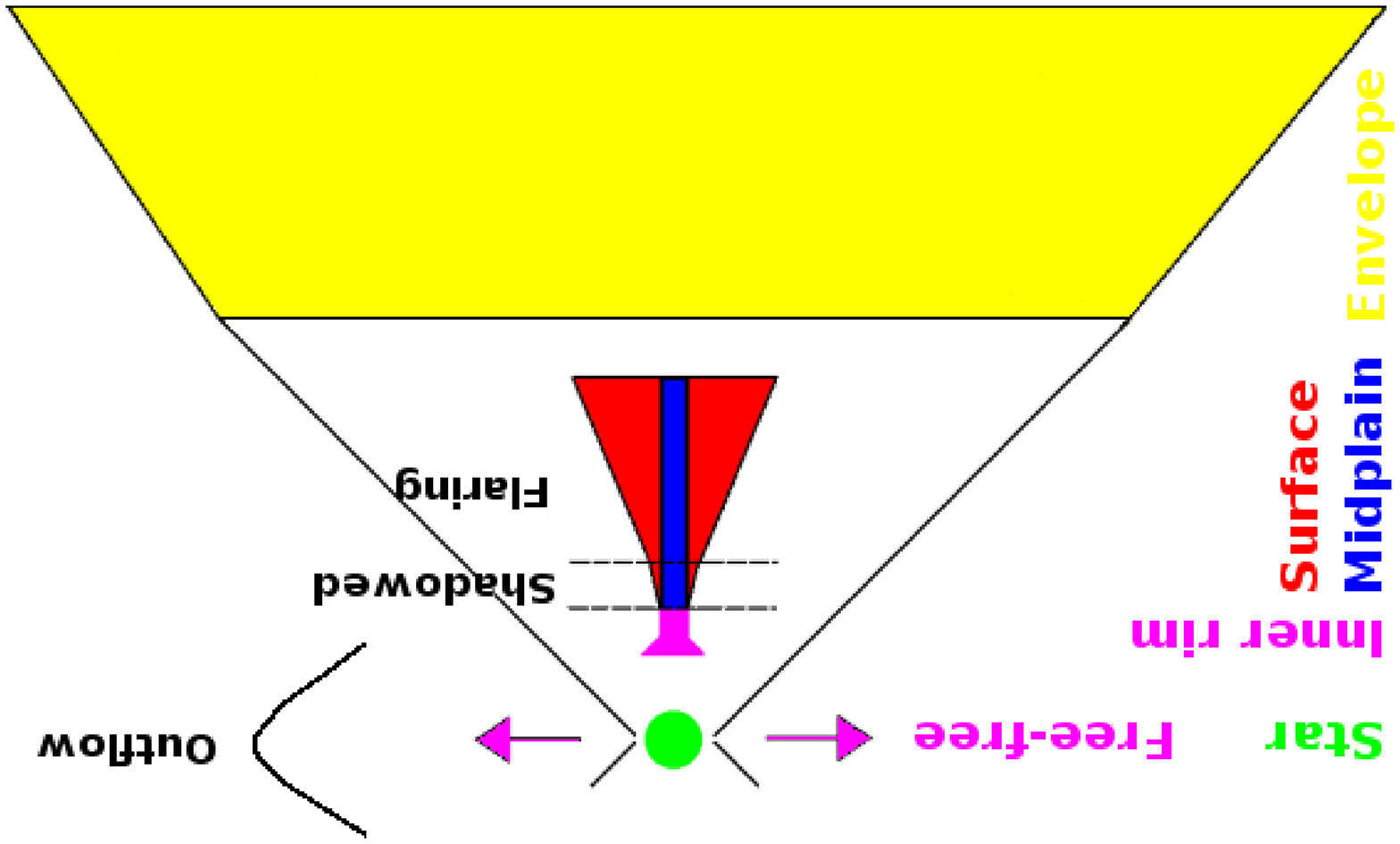}
\caption{
{\it Left:} Observed SED and model predictions for R~Mon. {\it Right:} Sketch of our model. The different emission components that contribute to the SED as predicted by our model are drawn in different colors. The emission of the star itself is drawn in green. The disk emission is separated into three components: inner rim emission (magenta), surface layer emission (red), and the midplane emission (blue). Note that the disk around R~Mon is fitted without the presence of an inner rim. The emission from the envelope is drawn in yellow. The free-free emission is shown in magenta. The disk emission and the disk+envelope emission appear as two continuous, black lines.}
\label{rmon_fit}
\end{figure*}

\section{Individual sources}
\label{individualSources}

\subsection{R Mon}

The SED towards R Mon is the only in our sample for which the first peak occurs at 4-5 $\mu$m instead of at 2 $\mu$m (see Fig. \ref{rmon_fit}). As a result, the SED can be fitted by a Chiang \& Goldreich-like model without the presence of an inner rim. However, an inner rim cannot be discarded because we lack of photometry between 2 $\mu$m and 3 $\mu$m, and extinction at these wavelengths is very high. The near and mid-IR part of the SED is well fitted by the emission of a circumstellar disk with r$_{out}$$\sim$150~AU. The interferometric 1.3~mm and 2.7~mm data allow us to constrain the mass of dust and estimate the grain sizes. The dust mass in the disk around R~Mon is found to be 1.4x10$^{-4}$~M$_\odot$ and grain growth has proceeded to sizes of $\sim$~1~cm in the midplane (see Table \ref{disks} and \cite{fue03,fue06}). \cite{fue06} and \cite{alo07} demostrated the existence of a gaseous, flat disk around R~Mon. The existence of large grains in the disk midplane is consistent with this geometry. Large grains produce a decrease in the dust opacity and a more rapid photoevaporation of the external layers of the gaseous disk.

The difference between the single-dish and interferometric fluxes at 1.3mm is clearly indicative of the presence of an envelope. \cite{clo97} and \cite{mov02} inferred that the nebula illuminated by R Mon has an hourglass shape, and an inclination angle of $\sim$20$^\circ$ with respect to the sky plane. The same kind of hourglass morphology is observed in the $^{12}$CO 1$\rightarrow$0 and 2$\rightarrow$1 images reported by \cite{fue06}. To model the envelope, we used a toroid with the same inclination as observed and shadowed by the inner disk. The available photometry includes Spitzer mid-IR observations with a beam size of 30$''$ and JCMT observations from \cite{man94} with a beam size of around 18$''$. We extended the envelope to cover the Spitzer beam size, using a temperature profile $T(r) = 62 (r(AU)/700)^{-0.5}$ K (see Table \ref{envelopes}). The mass of dust in the envelope was found to be 0.008~M$_\odot$.
The density profile of the envelope was described by a power law with slope -1.4. This means that an important fraction of the mass is close to the inner radius of the envelope, which is found to be 700 AU. The dust temperature at this radius is 60~K.


\subsection{Z~CMa}

\begin{figure*}
 \includegraphics[viewport=0 0 500 580, width=0.38\textwidth, height=0.5\textwidth, angle=-90]{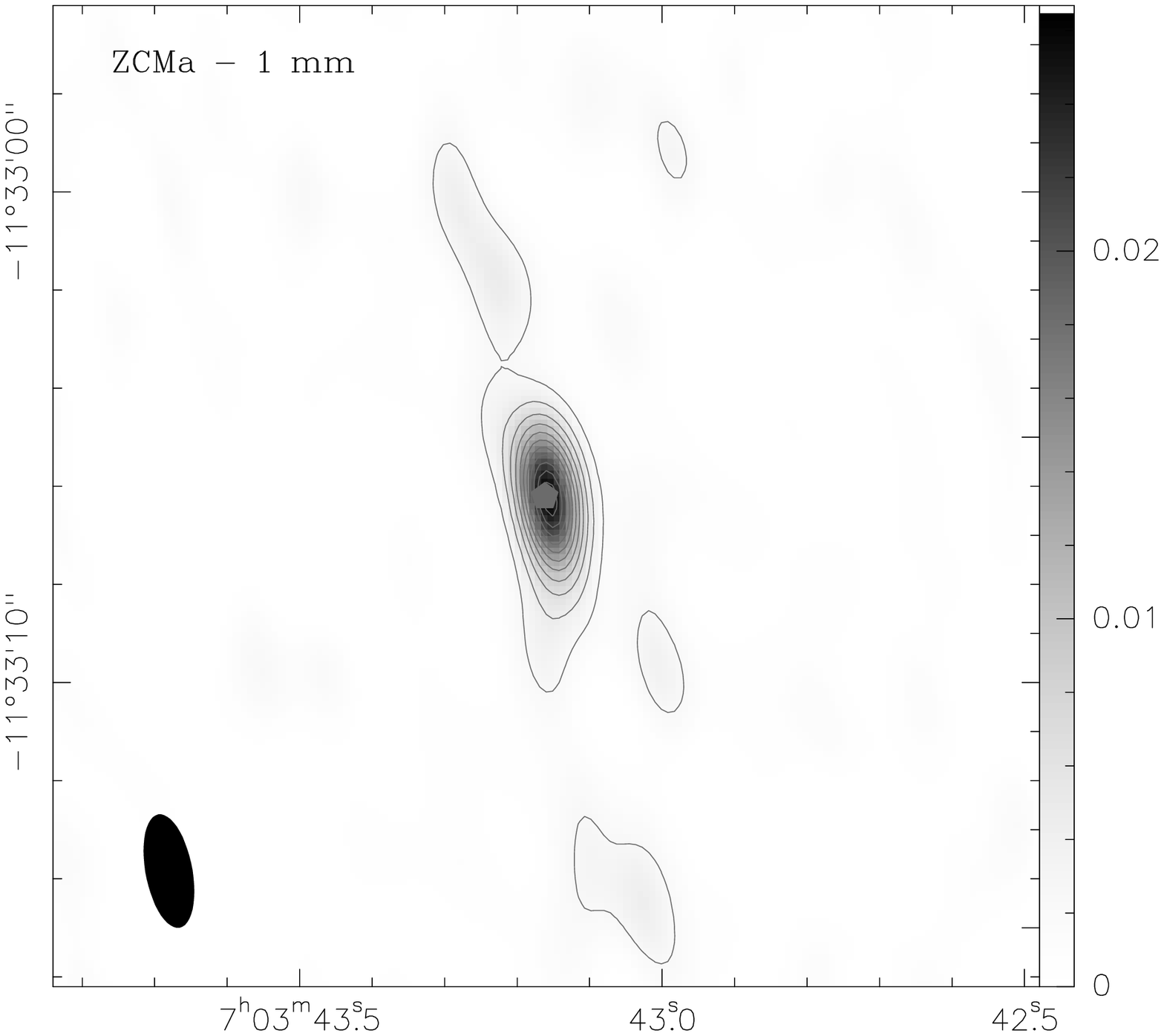}
 \includegraphics[viewport=0 0 500 580, width=0.38\textwidth, height=0.5\textwidth, angle=-90]{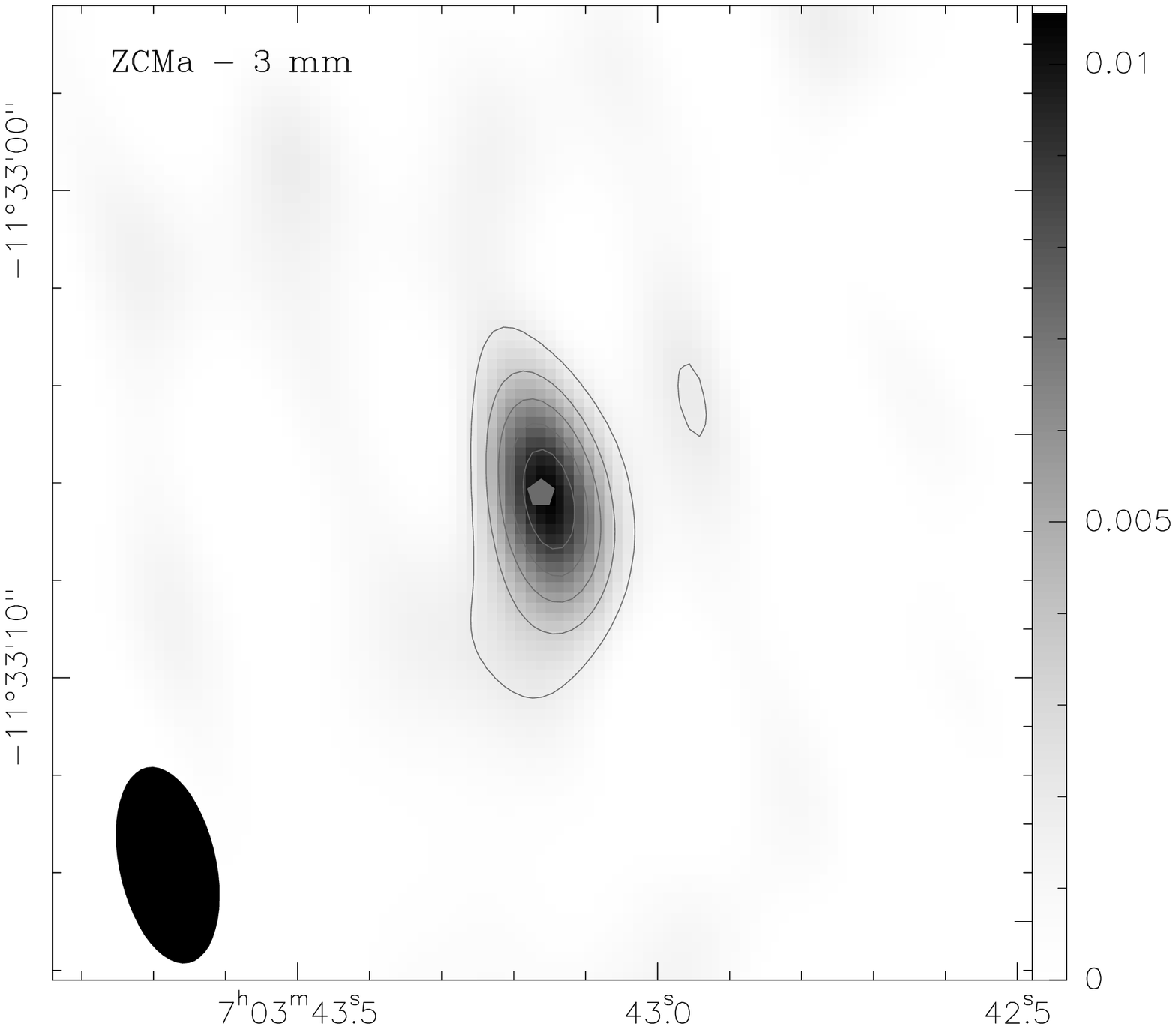}
 \includegraphics[viewport=-35 -65 465 475, width=0.4\textwidth, height=0.5\textwidth, angle=-90]{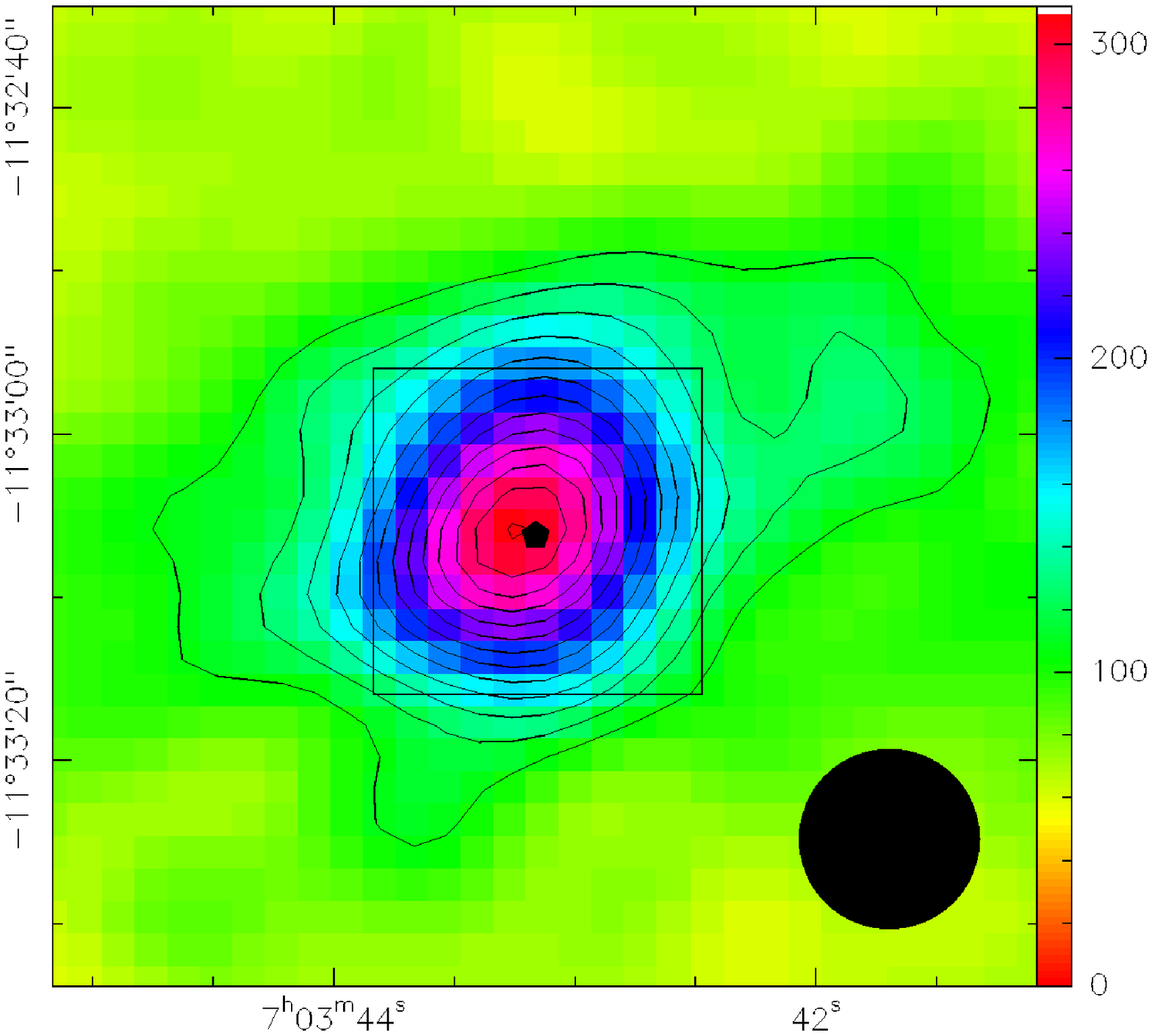}
 \includegraphics[viewport=-10 -30 500 700, width=0.4\textwidth, height=0.5\textwidth, angle=-90]{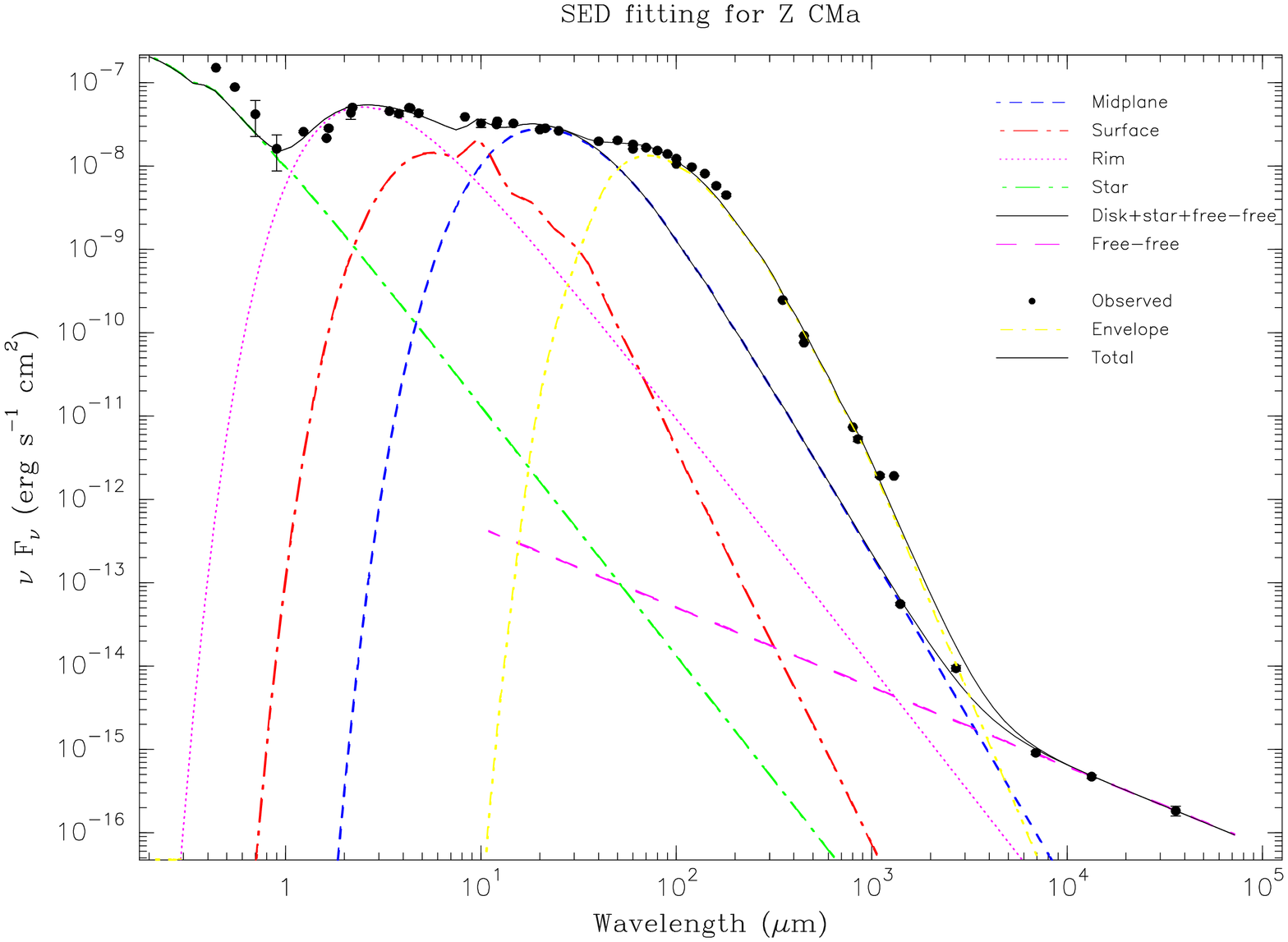}
\caption{{\it Up:} Z~CMa maps obtained with PdBI. Contours in the 1.3mm image are represented in steps of 0.77~mJy/beam, starting from 2.31~mJy/beam. In the 2.6mm image, the contours are from 1.8~mJy/beam, in steps of 0.6~mJy/beam. {\it Down-Left:} Z~CMa map at 1.3~mm obtained with the 30m bolometer. Contours are represented in steps of 20~mJy/beam starting from 50~mJy/beam (3$\sigma$ level). Maximum intensity is 293~mJy. The rectangle indicates the 20 arcseconds side region shown in the upper panels. {\it Down-Right:} The observed SED and our model predictions for Z~CMa. The different emission components are drawn following the color code described in Fig. \ref{rmon_fit}.
}
\label{zcma_fit}
\end{figure*}

Z~CMa is a binary system with a HBe star and a FU Orionis companion. The most recent estimate of the stellar masses of Z~CMa and its companion were 16~M$_\odot$ and 3~M$_\odot$, respectively \citep{van04}. The available photometry only provides the combined emission of both components, with the exception of the observations by \cite{kor91} which resolved the two stars. The optical observations shown in Fig. \ref{zcma_fit} should be considered to be a combination of both components. We fitted the optical observations by assuming that the HBe star contributes approximately 30$\%$ of the total flux \citep{van04}. 

The SED of Z CMa is shown in Fig \ref{zcma_fit}. The shape from near to far IR is flat with an abrupt slope in the emission beyond 100-200 $\mu$m. These two characteristics indicate that this is a massive, compact object. Obviously, the disk properties are difficult to derive and the existence of two circumstellar disks, one around each star, and/or a circumbinary disk cannot be discarded. We attempted to reproduce the NIR, mid-IR and FIR SED by assuming one single disk around the HBe star. We were unable to fit the IR SED by assuming a value of 16~M$_\odot$ for the stellar mass, but we obtained a reasonably good fit by assuming a stellar mass of 12~M$_\odot$ which is still consistent with being a Be star. On the basis of mainly our interferometric mm observations, we derive a mass of $\sim$7.0x10$^{-4}$~M$_\odot$ for the circumstellar dust. The 1.3mm/2.7mm spectral index was fitted most accurately with grains of maximum size $\sim$~1~cm. 

The envelope around Z~CMa is the most massive and compact in our sample (see Fig. \ref{zcma_fit} and Table \ref{envelopes}), which may be interpreted as proof of the youth of this object. It is well fitted by a shadowed toroid of $r_{in}$= 2000~AU, $r_{out}$=5000~AU, and an inclination angle of 30$^\circ$. The abrupt slope in the sub-mm region implies that the envelope cannot be extended beyond an outer radius of 5000~AU.
For this source, some measurements derived for different beams of fluxes at 450~$\mu$m, 850~$\mu$m, and 1.3~mm are available \citep{den98,san01}. By fitting the observed flux by different beam models, we inferred a density slope of -0.6 for the envelope, a value close to the density slope of -0.7 found in the disk.

The derived extinction from the envelope model (Table \ref{zcma_ext}) is 0 since the inclination of the toroid is small, 30$^\circ$, close to a pole-on view, and the star itself is visible through the conical cavity excavated by the outflow. However, by increasing the inclination to 50$^\circ$, the extinction increases to a value a factor of 10 higher than the observed value. An outer cold, spherical envelope could possibly account more accurately for the observed extinction. 

\subsection{MWC~297}
\label{discuss297}

\begin{figure*}
 \includegraphics[viewport=0 0 500 580, width=0.38\textwidth, height=0.5\textwidth, angle=-90]{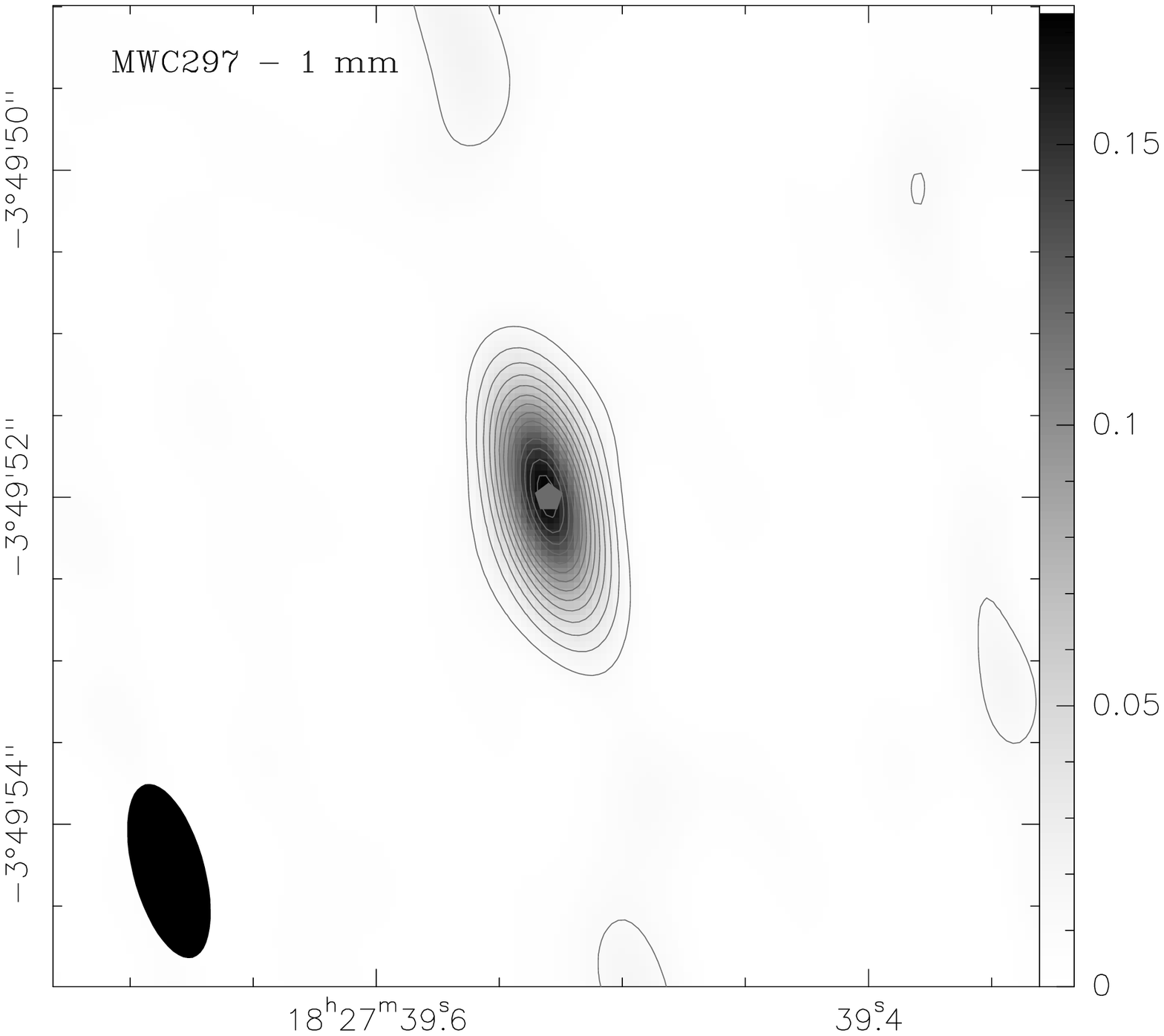}
 \includegraphics[viewport=0 0 500 580, width=0.38\textwidth, height=0.5\textwidth, angle=-90]{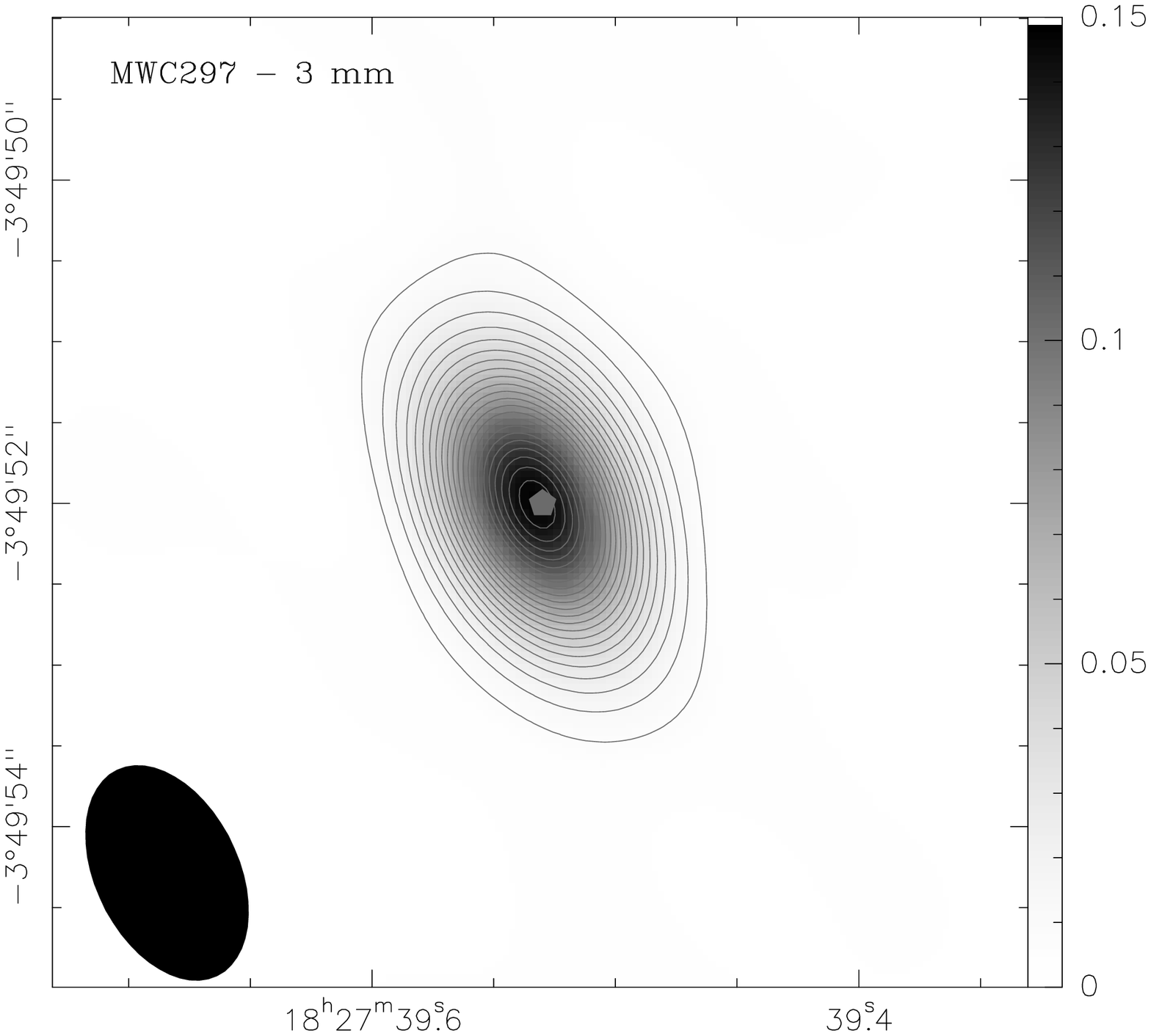}
 \includegraphics[viewport=-15 0 500 580, width=0.38\textwidth, height=0.5\textwidth, angle=-90]{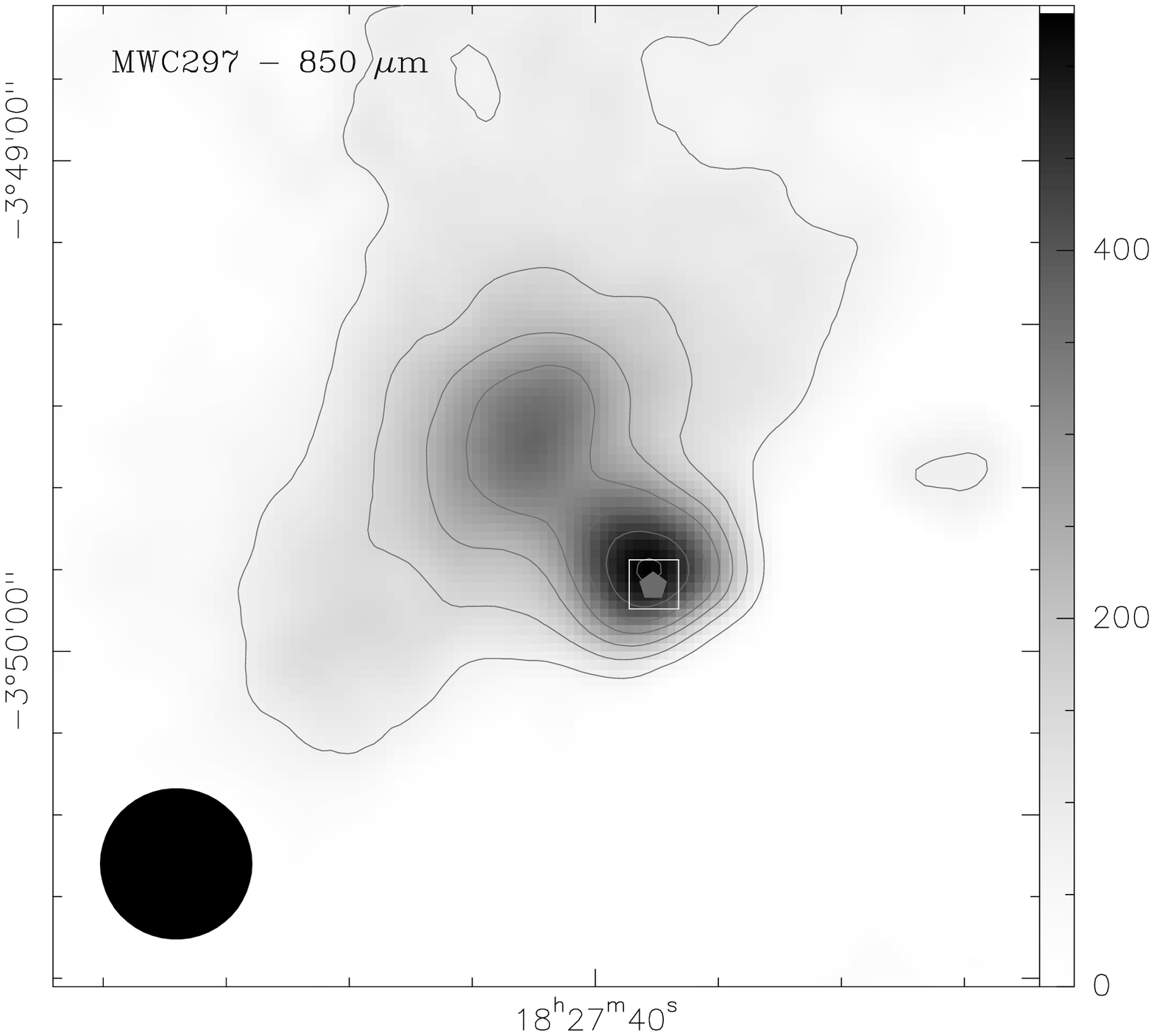}
 \includegraphics[viewport=-20 -30 500 700, width=0.4\textwidth, height=0.5\textwidth, angle=-90]{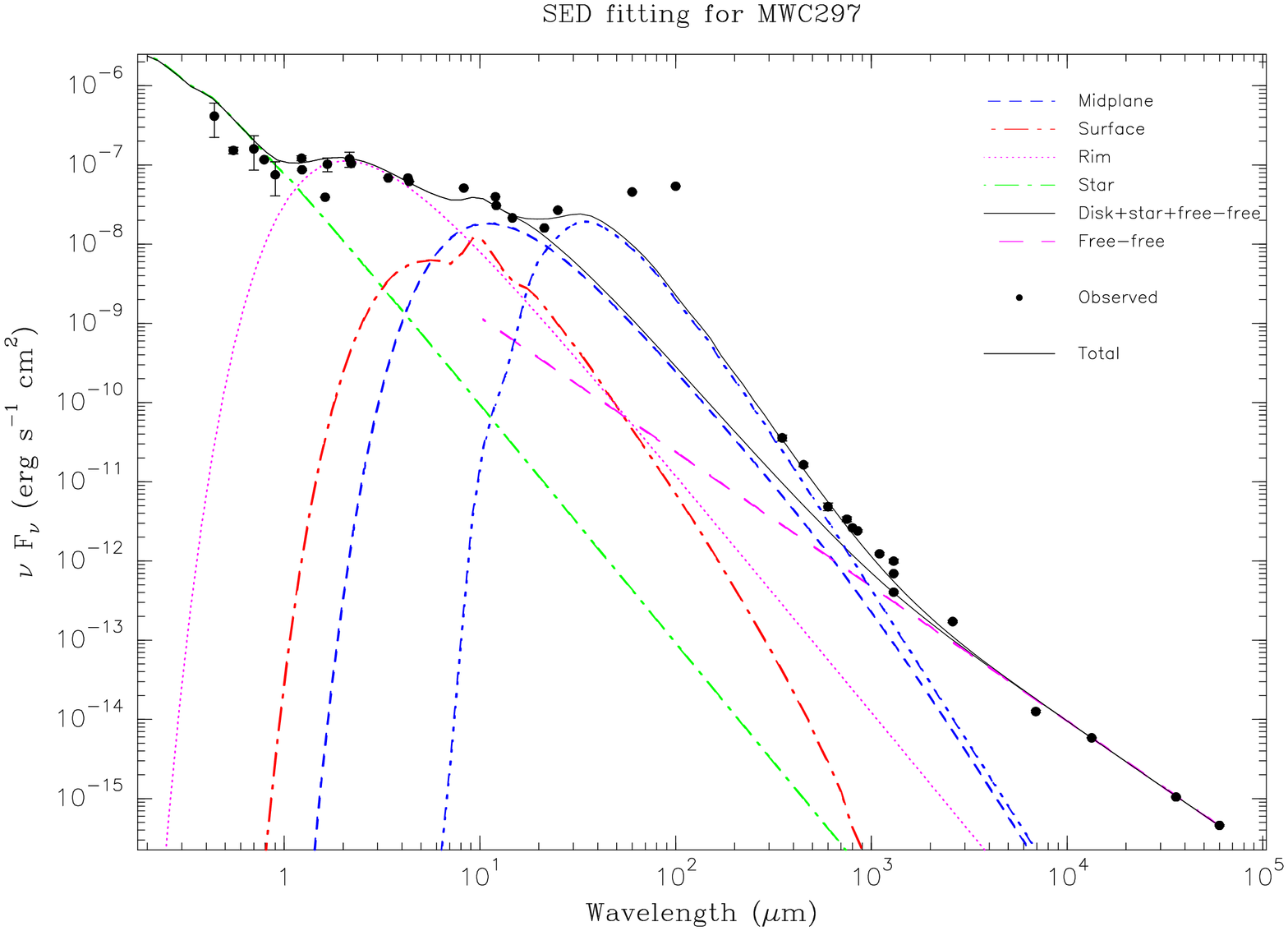}
\caption{{\it Up:} MWC~297 maps obtained with PdBI. Contours in the 1.3 mm image are represented in steps of 4~mJy/beam, starting from 12~mJy/beam. In the 2.6 mm image the contours are from 6.6 mJy/beam, in steps of 2.2 mJy/beam. {\it Down-Left:} MWC~297 map at 850 $\mu$m obtained with JCMT. Contours are represented in steps of 75~mJy/beam, starting from 75~mJy/beam. The small rectangle marks the 6 arcseconds side region shown in the upper panels. {\it Down-Right:} The observed SED and our model  predictions for MWC~297. Note that in this case there is no envelope but two disk components. 
}
\label{mwc297_fit}
\end{figure*}

At a distance of only 250~pc, MWC~297 is the closest star in our sample. It is highly reddened with a B-V color of 2.0, with no known binary companion. At optical wavelengths, a value of $R_v$ = 3.5 accounted for the extinction better than the value of $R_v$ = 3.1. At 2~$\mu$m, the SED is almost flat with a small prominence that could be interpreted as an inner rim (see Fig. \ref{mwc297_fit}). However, this feature is unclear. We obtained reasonably good fits by using a face-on disk ($i \sim 5^\circ$) model with an inner rim, or a highly inclined disk  ($i \sim 80^\circ$) without an inner rim. On the basis of IR interferometric observations, \cite{ack08} proposed the existence of a low inclination disk  
($i$ less than $40^\circ$) with dust closer to the star than the dust sublimation radius. We selected the low-inclination solution to be consistent with previous near-IR interferometric measurements by \cite{ack08}. Since our model is based on hydrostatic equilibrium, it cannot account for the existence of dust closer to the star than the sublimation radius. In our model, we allow the rim temperature (T$_{rim}$) to vary between
1500~K (silicates sublimation temperature) and 2000~K (graphite sublimation temperature) and the inner radius is determined by T$_{rim}$. The fact that the flux decreases at wavelengths longer than 10 $\mu$m implies that the circumstellar disk should be small. The most consistent fit is obtained with T$_{rim}$$\sim$1800~K, an outer radius of only $\sim$28~AU, a density slope of $\sim$-0.77, 
and an inclination angle of 5$^\circ$ $\pm$ 5$^\circ$ . This small disk is optically thick at 1.3mm and, consequently, 
the dust mass and grain properties are not determined (see Table \ref{grains}). We  assume $a_{max}$ = 1~cm in the midplane.


The slope of the SED at mm and sub-mm wavelengths is inconsistent with that expected from an envelope with standard interstellar-medium (ISM) grains. 
The 1.3~mm/2.7~mm spectral index is also too low to be adjusted by thermal dust emission. The problem could be the different beam sizes of the interferometric observations at 1.3~mm and 2.7~mm. In the case that the emission at 1.3~mm and 2.7~mm is more extended than the beams, the measured 1.3~mm/2.7~mm spectral index would be affected by the different angular resolutions. To confirm this possibility, we compared our 1.3~mm flux with the previous measurement carried out with the SMA providing a beam of $\sim$3$''$ \citep{man07}. Our flux 58\% of the flux values measured by \cite{man07}, indicating that our 1.3mm observations resolve the 1.3mm emission. Thus, we adopted the flux measured by \cite{man07} to derive the 1.3mm/2.7mm spectral index. Even using this value, the derived spectral index was too low to be due to the ISM grains expected in the envelope. The spectral index was, however, more similar to that found in circumstellar disks, where grain growth occurred. This prompted us to consider the existence of a second disk component towards MWC~297. This second disk could be due to a binary companion, or a ring around the main disk. Since no binary companion was detected close to this star (at a distance of $\sim$ 0.7\arcsec or 180 AU), the second option is more likely. We found good agreement with observations by assuming a ring of large grains ($\sim$~1cm), a dust mass of 4x10$^{-4}$ M$_\odot$, an inner and outer radius of 200 AU and 300 AU respectively, and a temperature profile $T(r) = 125 (r(AU)/200)^{-0.5}$ K.

The JCMT map (bottom-left panel in Fig. \ref{mwc297_fit}) indicates that, in addition to the small disk and ring, there is a more extended envelope surrounding this object. Nebulous emission is also observed in the MIPS 24~$\mu$m image, suggesting that the star is still embedded in the molecular cloud. However, most of the emission at millimeter wavelengths (92\%) indicates a small region ($\sim$3$''$) around the star (see Appendix A.7). This is consistent with our two-disk-component fit.

\subsection{MWC~1080}

MWC~1080 was classified as a B0 star. Its distance estimates ranged from 1000~pc \citep{eis04, fue03} to 2100~pc \citep{ack05}. We adopted the distance of 1000~pc, since it is far more consistent with the observed stellar emission after the extinction correction. 

\begin{figure*}
 \includegraphics[viewport=0 0 500 580, width=0.38\textwidth, height=0.5\textwidth, angle=-90]{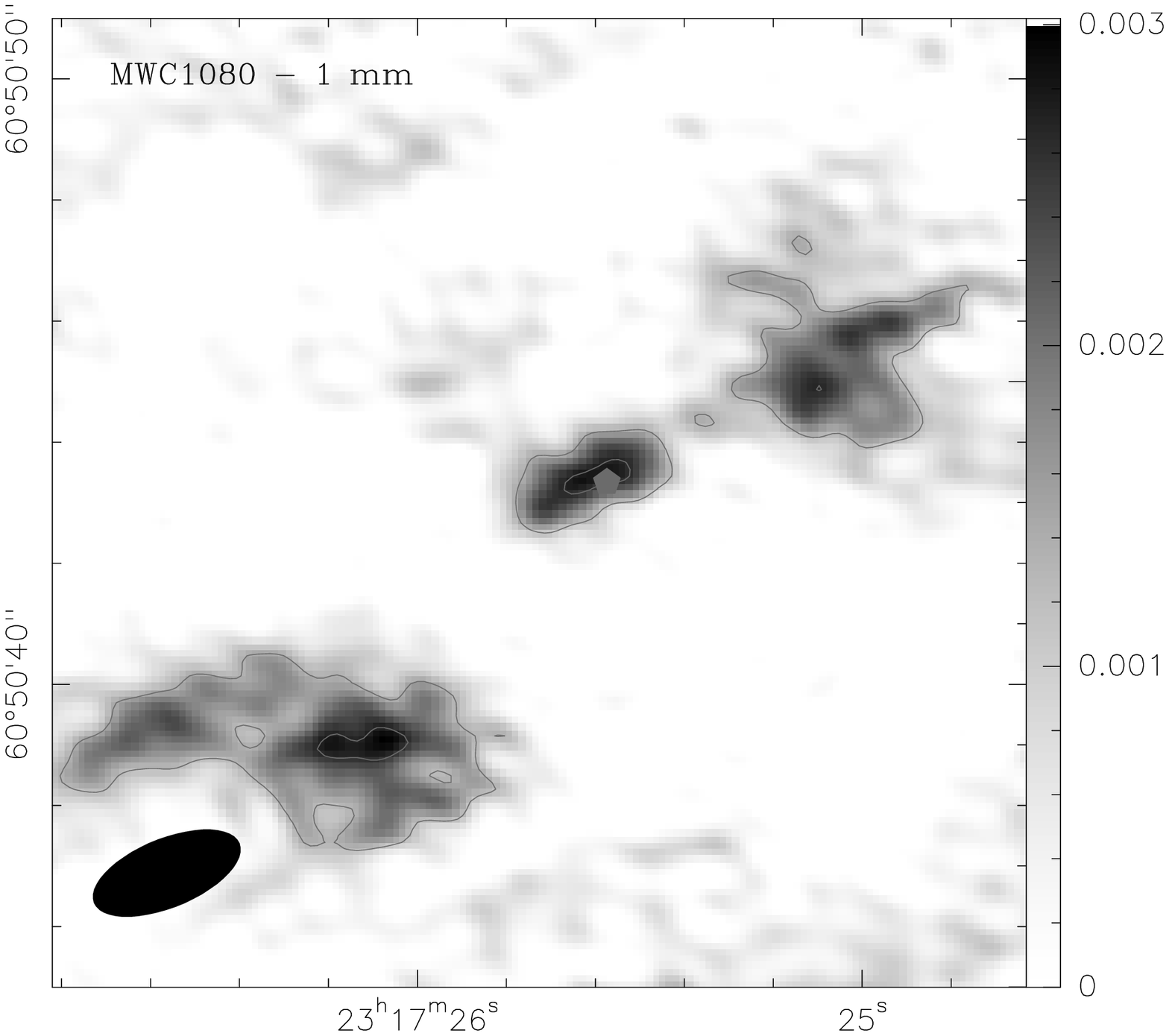}
 \includegraphics[viewport=0 0 500 580, width=0.38\textwidth, height=0.5\textwidth, angle=-90]{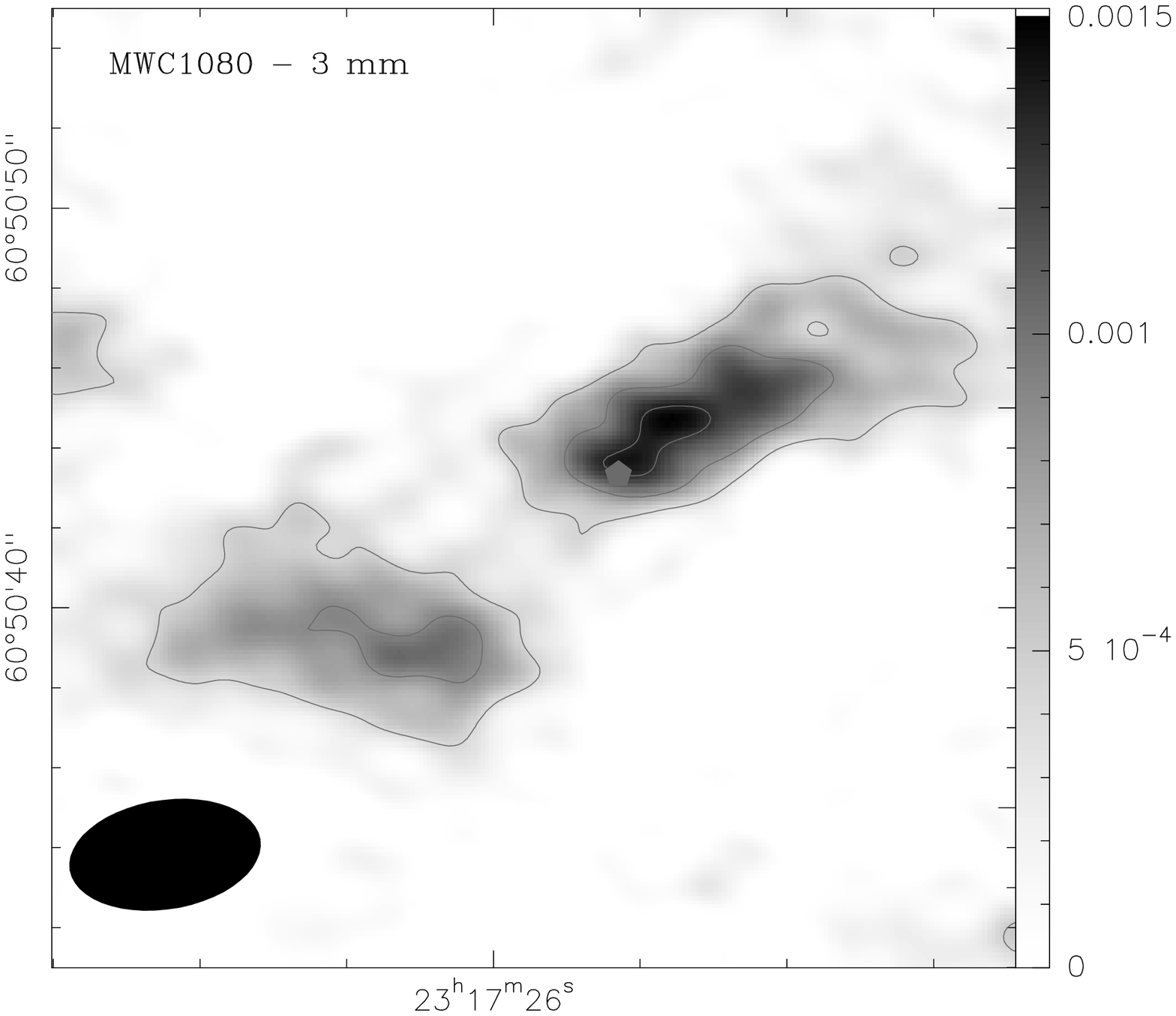}
 \includegraphics[viewport=-35 -65 465 475, width=0.4\textwidth, height=0.5\textwidth, angle=-90]{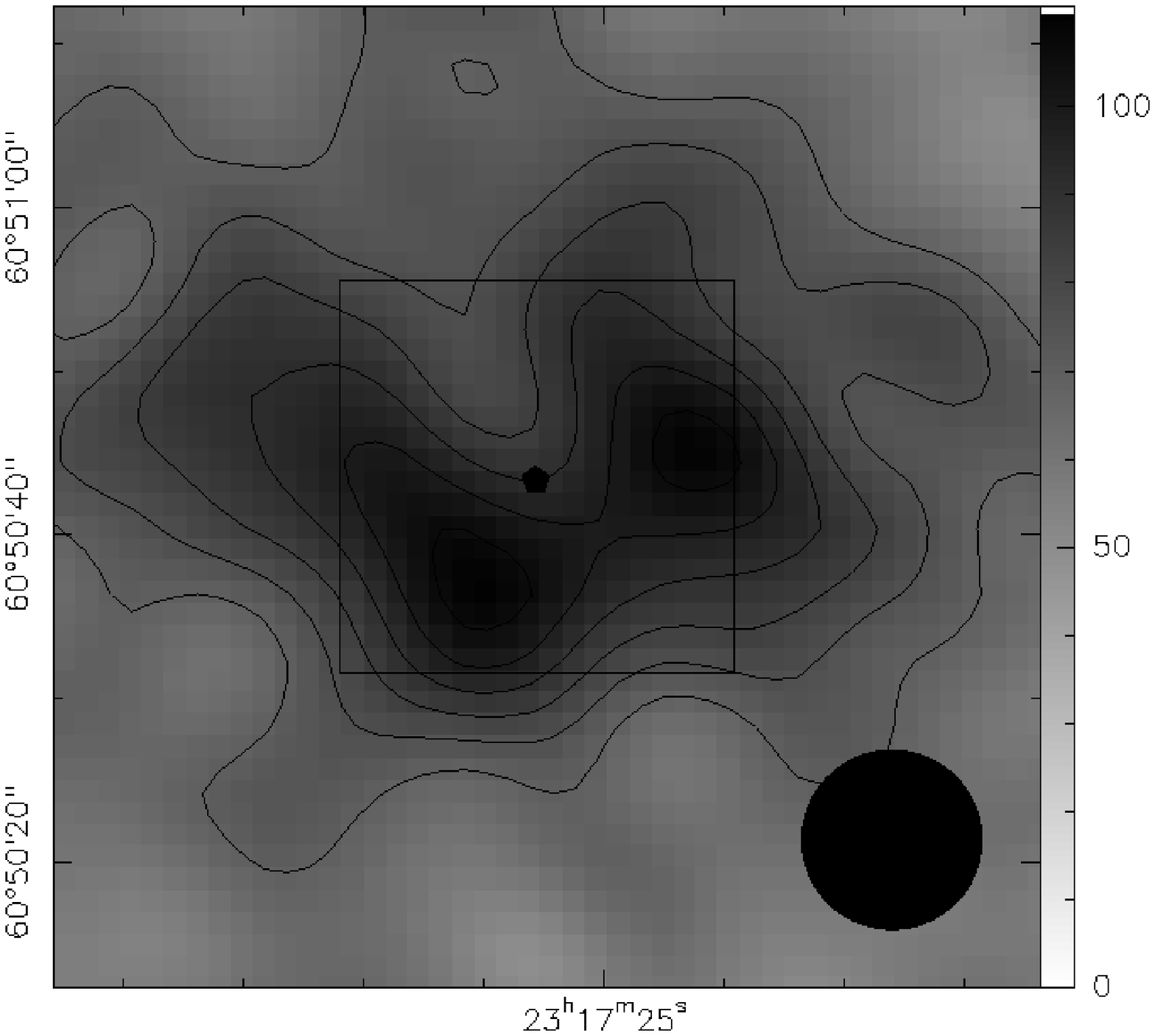}
 \includegraphics[viewport=-10 -30 500 700, width=0.4\textwidth, height=0.5\textwidth, angle=-90]{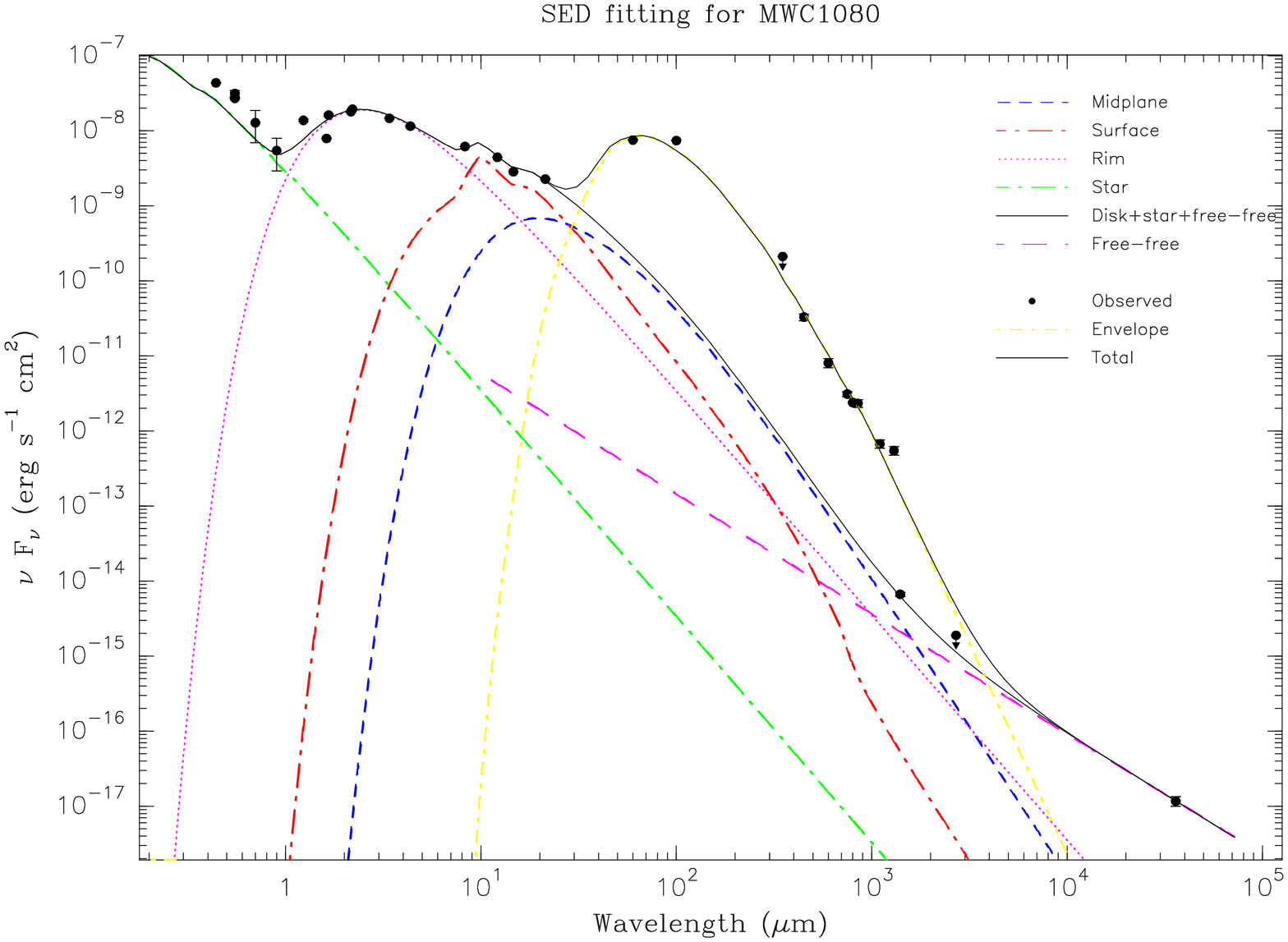}
\caption{{\it Up:} MWC~1080 maps at 1.3 mm and 2.6 mm obtained with PdBI. Contours in the 1.3 mm image are represented in steps of 0.45 mJy/beam, starting from 1.35 mJy/beam. In the 2.6 mm image, the contours are from 0.45 mJy/beam, in steps of 0.15 mJy/beam. {\it Down-Left:} MWC~1080 map at 1~mm obtained with the 30m bolometer. Contours are represented in steps of 10~mJy/beam starting from 40~mJy/beam (3$\sigma$ level). Maximum intensity is 95~mJy. The 26 arcseconds side square indicates the region shown in the PdBI 2.6mm map. {\it Down-Right:} The observed SED and our model predictions for MWC~1080. }
\label{mwc1080_fit}
\end{figure*}

The SED towards MWC~1080 has a clear bump at 2 $\mu$m, indicating the presence of a rim with T$_d$$\sim$1500~K (see Fig. \ref{mwc1080_fit}). The dust temperature in the inner rim was determined from the peak at 2 $\mu$m. The mid-IR part of the SED was well fitted with a disk of r$_{out}$$\sim$77~AU and an inclination angle of $\sim$80$^\circ$. This inclination angle is totally consistent with MWC~1080 being an eclipsing binary star, for which the amplitude in the variation of the apparent magnitude is wide \citep{gra92}. The dust mass that better accounts for the 1.3~mm interferometric flux is found to be 5x10$^{-5}$~M$_\odot$. The maximum grain size is not determined since we have not detected the disk at 2.7mm. We obtain a good fit for grains of $\sim$1~cm in the midplane.

The existence of an envelope is certain since the observed 1.3mm interferometric flux is a small fraction (1.3\%) of the total flux measured with the JCMT. The interferometric 1.3 and 2.7mm images also reveal the existence of several clumps surrounding the circumstellar disk (see Fig. \ref{mwc1080_fit} and Fuente et al. 2003). On the basis of the morphology observed in our interferometric images, we used a toroid with an inner radius of 6000~AU (6$''$ at the distance of MWC~1080) to fit the envelope. The best fit solution for the envelope is a non-shadowed toroid of an outer radius of 12 000~AU (12$''$). Our model fits all the SED points except the IRAS flux at 100 $\mu$m. This is unsurprising since the wide IRAS beam at 100~$\mu$m ($\sim$2$'$) probably detects emission from the foreground molecular cloud.

The extinction estimate from the B-V color index is 40$\%$ higher than that predicted by our envelope model. This is also consistent with the scheme of the star+disk+toroid system immersed in a lower-density molecular cloud.

\subsection{MWC~137}

\begin{figure}
 \includegraphics[viewport=0 10 500 840, width=0.45\textwidth, height=0.45\textheight, angle=-90]{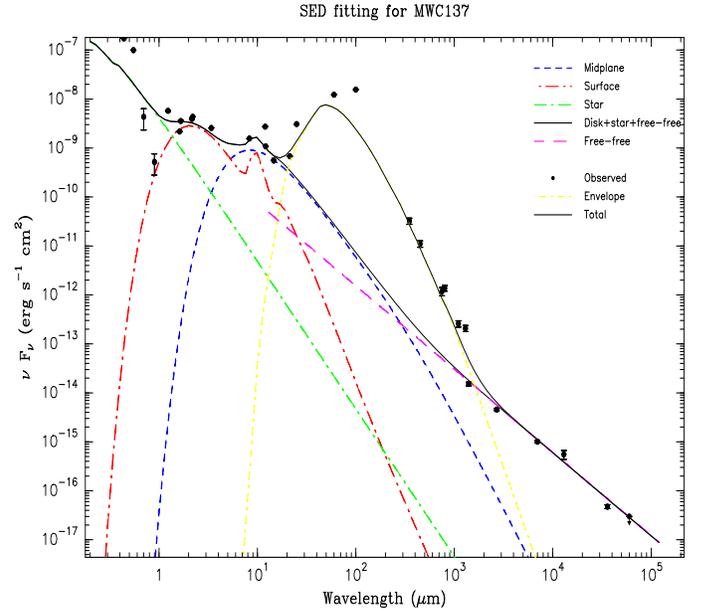}
\caption{
The observed SED and our model predictions for MWC~137.}
\label{mwc137_fit}
\end{figure}

The SED of MWC~137 does not exhibit any evidence of disk emission at mm wavelengths. 
All the mm-cm fluxes can be fitted
by a single component of spectral index $\alpha$ = +0.76$\pm$0.01. 
This spectral index is consistent with that expected in the free-free emission 
arising in the stellar wind. Although a value of $\alpha$ =+0.6 is expected
for an ionized isotropic wind, small deviations of this value can be explained by
a different geometry or a partially ionized wind. However, 
some excess is visible at IR wavelengths, which is suggestive of the presence of a small disk. 
\cite{fue03} derived an upper limit to the dust mass of 7x10$^{-5}$~M$_\odot$ assuming that 
the disk is optically thin at 1.3mm, the free-free emission spectral index is +0.6, and standard 
values for the dust opacity and mean dust temperature (see Sect. 6). 
However, this value is quite uncertain since the disk is probably optically thick and the 
grain emissivity is unknown. 
We fitted the NIR part of the SED and obtained the best-fit solution (consistent with our mm observations) 
for a small disk of r$_{out}$$\sim$18~AU and a dust mass of 10$^{-5}$~M$_\odot$. The absence of a large 
disk ensure that the envelope is prominent in the SED. The envelope is modeled as an sphere 
extended between 5000 and 11 700 AU, with a dust mass of 0.005~M$_\odot$ (see Fig. \ref{mwc137_fit}).


\begin{figure}
 \includegraphics[viewport=0 10 500 840, width=0.45\textwidth, height=0.45\textheight, angle=-90]{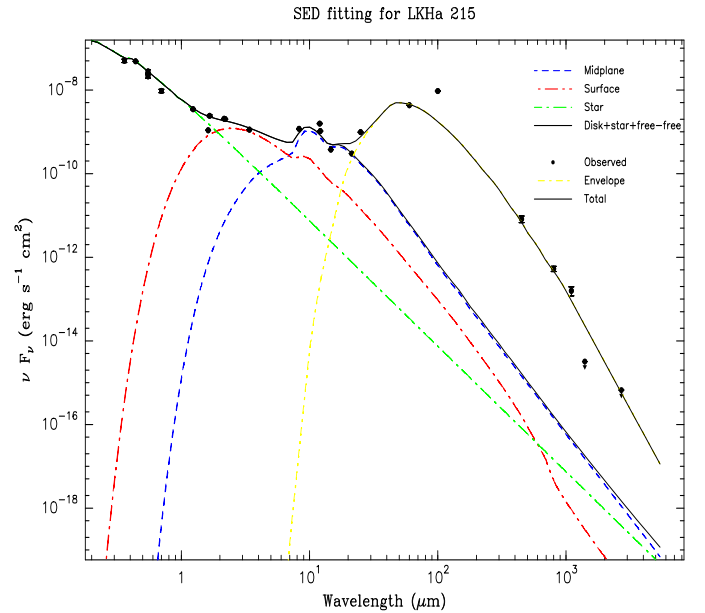}
\caption{The observed SED and our model predictions for LKH$\alpha$~215 .
}
\label{lkha215_fit}
\end{figure}

\subsection{LKH$\alpha$~215}

We did not detected this source at mm and cm wavelengths. This implies an upper limit to the dust mass of 9.0x10$^{-5}$~M$_\odot$, assuming optically thin emission and standard values of the dust temperature and emissivity. We modeled the entire SED to improve the upper limit to the mass of dust and to determine the envelope component (see Fig. \ref{lkha215_fit}). Our best fit, still although uncertain, solution is obtained for a disk with an outer radius of $\sim$10~AU and a mass $\sim$6x10$^{-8}$~M$_\odot$. The envelope is described well by a dust mass of about $\sim$0.0015 M$_\odot$, located between $\sim$3500~AU and $\sim$7200~AU.

\begin{figure*}
 \includegraphics[viewport=0 10 500 720, width=0.45\textwidth, height=0.38\textheight, angle=-90]{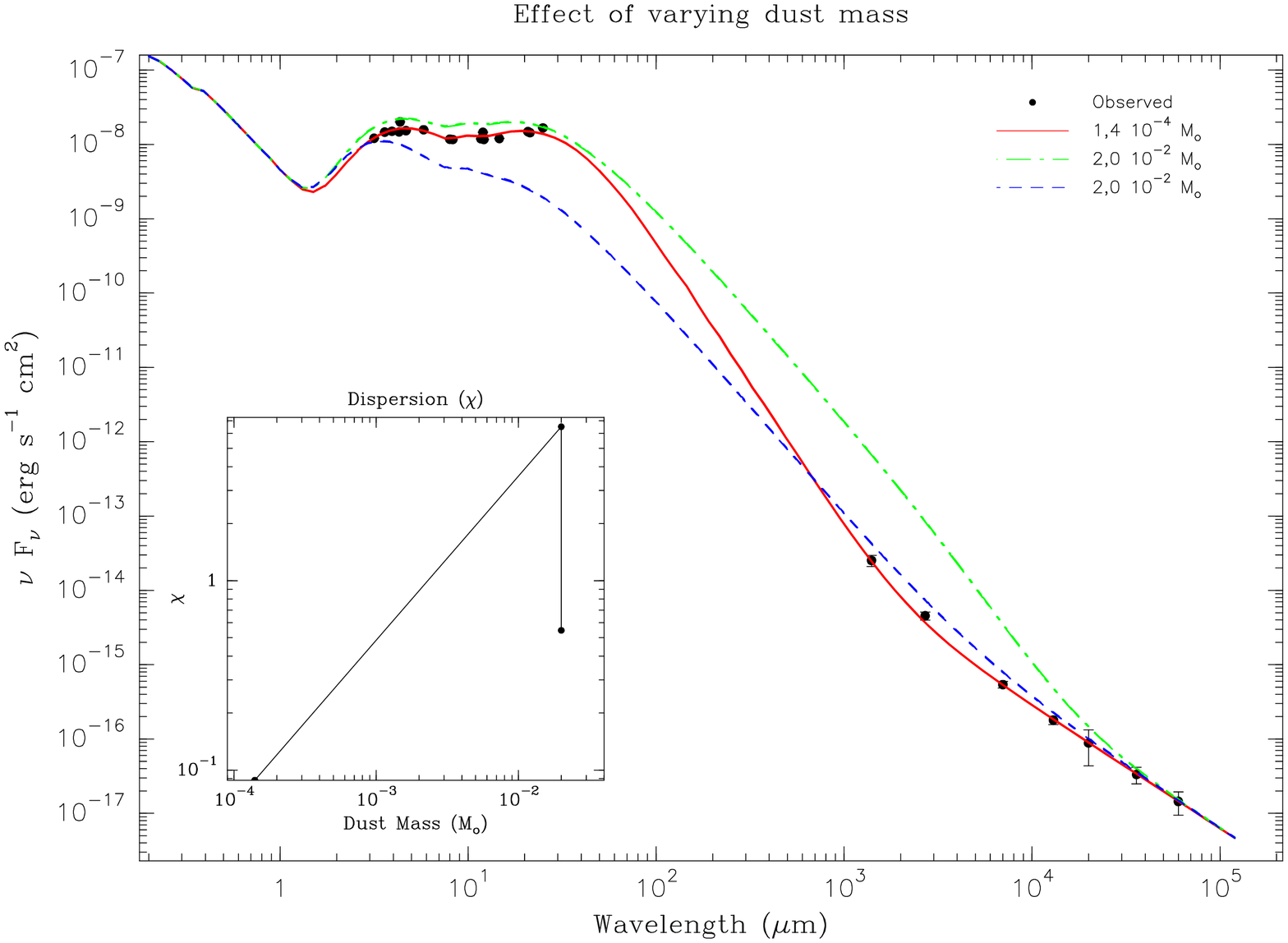}
 \includegraphics[viewport=0 10 500 720, width=0.45\textwidth, height=0.38\textheight, angle=-90]{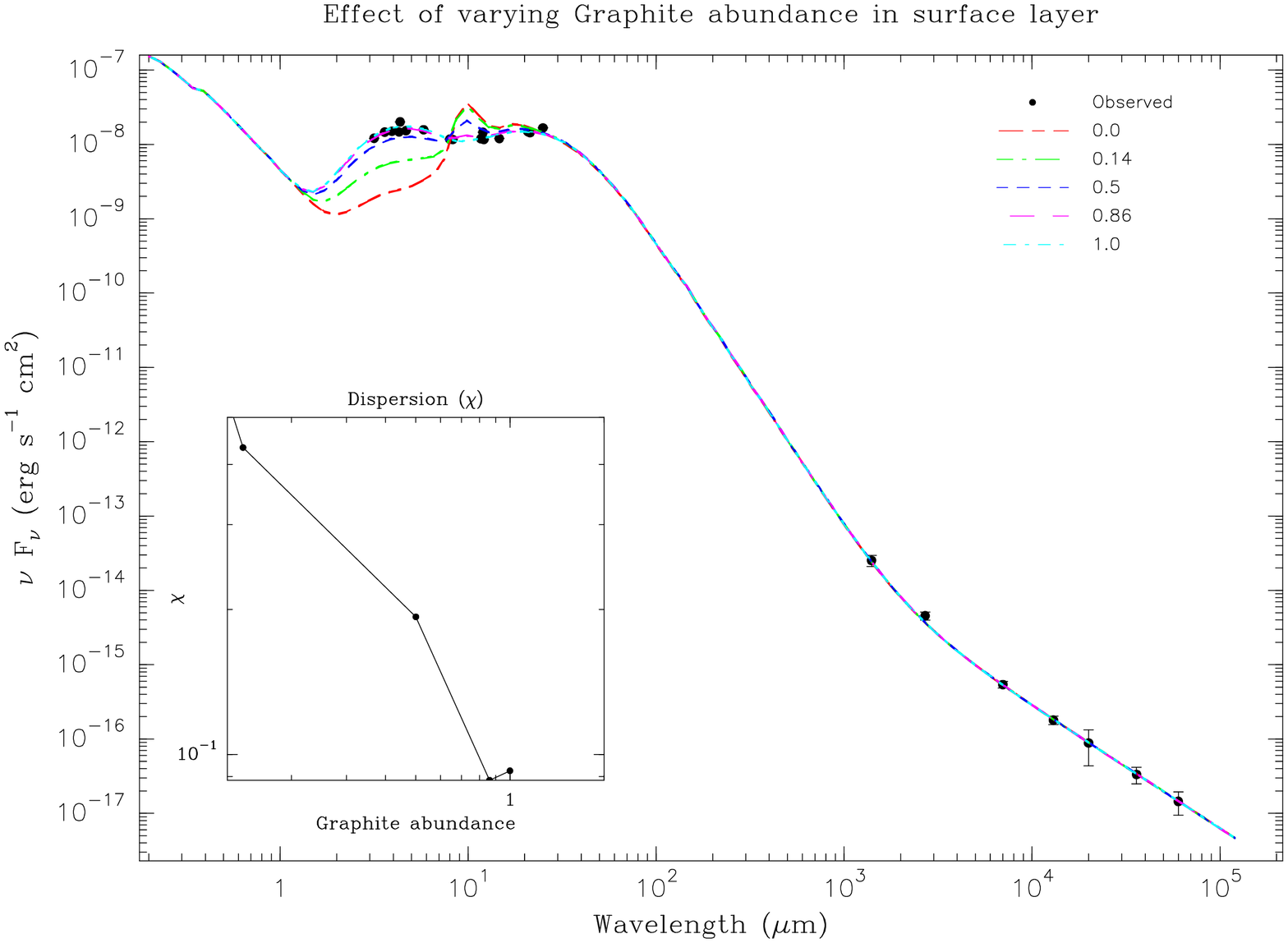}
\caption{{\it Left:} The SED of R~Mon cannot be fitted assuming an optically thick disk. In this case, we obtain a large excess at mm wavelengths (green short-long dashed line). To reduce this excess, we can adjust the other disk parameters, such as the inclination angle or the outer radius, to reduce the effective surface of the disk. But in this case, we obtain a deficient emission at near and mid-IR wavelengths (blue dashed line). The red continuous line shows our optically thin solution. The lower-left chart shows the evolution in the dispersion parameter ($\chi$). {\it Right:} The SED of R~Mon and Z~CMa cannot be fitted using a standard grain mixture (86\% of astronomical silicate and 14\% of graphite). The pink line shows our best fit for a non-standard mixture with 86\% of graphite. With 100\% of graphite the fit is still excellent, but slightly worse. When the fraction of silicates becomes important (50\% or higher) the model is inconsistent with the observations.}
\label{rmon_mass_fit}
\end{figure*}


\subsection{Uncertainties in the disk mass and maximum grain size}

The dust mass and the grain properties determined by fitting the SED are affected by important uncertainties. The most important assumption is that the dust emission is optically thin at mm wavelengths. This is the case for typical circumstellar disks of $r_{out} \sim 100$~AU and dust mass $< 10^{-3}~M_\odot$. Only in the case of small disks, the 1.3mm emission could become optically thick. 

We discuss R~Mon as a representative case of an optically thin disk. Our best fit solution is for a disk with r$_{out}$=150~AU and a dust mass of 1.4x10$^{-4}$~M$_\odot$. This disk is optically thin at 1.3mm. A disk of this size needs to have a dust mass of $\sim$0.01~M$_\odot$, i.e. a factor of 100 higher than our estimate, to become optically thick ($\tau_{1.3mm} \sim$1). But in this case, the flux at 1.3mm would be more than one order of magnitude higher than the measured flux (see Fig. \ref{rmon_mass_fit}).  In the optically thick case, the flux at 1.3mm depends only on the area of the disk projected on the plane of the sky and the dust temperature across the disk surface, $F_{1.3mm} \sim B_{\lambda}(T_s) \pi r_{out}^2 cos(i)$. The only way of reconciling the predicted mm flux with observations is by reducing the size of the disk and/or changing the disk inclination. But in that case we obtain a deficient emission in the NIR and mid-IR range as it is illustrated in Fig. \ref{rmon_mass_fit}. The same argument is valid for Z~CMa.

A different case is the small disk surrounding MWC~297. In this case, the disk is optically thick and the derived mass is only a lower limit to the true disk mass. However, it is unreasonable to propose that the dust mass in the small and, probably, evolved disk around MWC~297 is larger than that within the less evolved disk around R~Mon. One would expect that the mass of dust and gas in the disk decreases because of photoevaporation (see Sect. 6). Thus, our mass should be accurate to within an order of magnitude.

The SED towards MWC~1080 is well described by a partially optically thin disk as shown in Table \ref{disks}. But it can also be reasonably well reproduced by an optically thick disk with r$_{out}$=50~AU and a dust mass $\geq$~10$^{-3}$ M$_\odot$. Following the same argument as for MWC~297, it is unreasonable to assume that the dust mass in this
small disk is one order of magnitude larger than in the less evolved disk around R~Mon and similar to those found in TTs. Thus, we keep the optically thin solution, although we are aware of the uncertainty in our analysis.

In the case of optically thin disks, we can learn about the grain properties from the 1.3mm/2.7mm spectral index. In the Rayleigh-Jeans region of the spectrum, the dust emission is $\propto$ $\nu^{(2+\beta)}$ where $\beta$ is the opacity index that can be accurately derived by fitting the observed SED. The value of $\beta$ in the submm-mm range is an excellent indicator of the grain size distribution \citep{dra06}. The grain size distribution is usually taken to be, $n(a)=n_0 a^{-p}$ for $a < a_{max}$, where n$_0$ is a normalization factor, $p$ is a parameter in the range 2.5--3.5, and $a_{max}$ is the maximum grain size. The value of $\beta$ is between 1 and 2 for small grains (sub-micron-sized grains) but becomes lower than 1 when $a_{max}$ is a few mm. To have a value of  $\beta$$\leq$0.5, we need grains with $a_{max}$$\sim$~1~cm. This is the case for R~Mon and Z~CMa. However, there is some degeneracy between the index of the grain size distribution, p, and the value of the maximum grain size, $a_{max}$. A slope p=2.5 means that a larger number of grains are closer to the maximum grain size than for a more typical slope of 3.5. By assuming p=2.5 we would reach the same values of $\beta$ with grains slightly smaller. Even in this case, we would require grains with radii of nearly 1~cm, i.e. significant grain growth must have occurred in the disk midplane to have the values of $\beta$ $<$0.5 that we obtain in our sources.  We cannot determine grain sizes larger than a few cm on the basis of mm observations. The 1.3mm/2.7mm grain opacity index tends asymptotically to 0 for $a_{max}~>$~a~few~cm. In the case of optically thick disks, we cannot determine the grain size from the SED. We assumed $a_{max}~$$\sim$1~cm in the midplane for MWC~1080, MWC~297, MWC~137, and LkH$\alpha$~215, since grain growth is expected to have proceeded in these more evolved disks.

The dust composition in the midplane cannot be inferred from the SED. For this reason, we assumed the standard mixture in all of our sources (see Table \ref{grains}). The silicate feature at $\sim$9.8 $\mu$m provides, however, some information about the dust composition in the surface layer. In all of our sources, the silicate feature is weak or absent. The absence of this feature does not imply the absence of silicate grains but the lack of silicate grains at a temperature of $\sim$800~K.
The silicate grains could be either too far from the central source and thus not heated to temperatures sufficiently high to allow emission in the mid-IR, or these grains could be too large to be heated efficiently. 
To investigate the second possibility, we tried to fit SEDs by assuming a standard dust composition and varying the $a_{max}$ in the surface between 0.1$\mu$m up to 100~$\mu$m. We were able to reproduce the disks around MWC~137, MWC~1080, MWC~297, and LkH$\alpha$~215 (see Table \ref{grains}), but failed to account for R~Mon and Z~CMa. In these cases, we needed to vary the dust composition to fit the mid-IR part of the SED. The most extreme case was R~Mon in which the best fit solution was found for 86\% of graphite in the surface layer (see Fig. \ref{rmon_mass_fit}). This result is, of course, limited by the simplicity of our model. A different disk geometry is expected in a thermal equilibrium disk. Alternatively, the dust composition and/or grain growth could vary as a function of the distance from the star. Determining the dust composition of the disk surface is, however, beyond the scope of this Paper, and, in addition, does not affect our results because the disk mass depends only on the dust composition and grain size in the midplane. Therefore, we do not discuss this point further.

The dust composition and maximum grain size in the midplane determine the value of dust emissivity at 1.3mm, $\kappa_{1.3mm}$, and consequently has an influence on the derived dust mass. Assuming a standard dust mixture and $a_{max}$=1~cm, $\kappa_{1.3mm}$=0.88~g$^{-1}$~cm$^{2}$. This value is close to the canonical value of 1~g$^{-1}$ cm$^{2}$.  For reasonable values of a$_{max}$ in disks, i.e. of between 1~mm and 1~cm, the value of $\kappa_{1.3mm}$ varies by a factor of 2--3 from the canonical value. This is the uncertainty in the disk mass estimate due to the uncertainty in the grain size in the midplane.

Summarizing we conclude that our disk-mass estimates are accurate to within a factor of 2--3 in the optically thin disks around R~Mon and Z~CMa, and to within an order of magnitude for the optically thick disks around MWC~1080 and MWC~297.


NIR interferometric observations by \cite{eis07} and \cite{ack08} revealed the existence of hot dust in a region at $<$~1~AU from the star towards MWC~297 and MWC~1080. 
We are aware that our model is too simple to determine the properties of the disk in a few~AU around the star. First of all, in our model, the inner radius of the disk is determined by the dust sublimation temperature that we assume to be between 1500~K (silicate sublimation temperature) and 2000~K (graphite sublimation temperature). We do not consider the possibility of the existence of dust closer to the star, as proposed by \cite{eis07} and \cite{ack08}. Secondly, this extremely hot dust (at less than 1~AU) would emit radiation at $<$2~$\mu$m and the SED at these wavelengths is dominated by the emission of the stellar photosphere. Our model and method (SED fitting) are indeed inadequate to determine the properties of the innermost regions of the disk (r$\leq$a few~AU). NIR interferometric observations capable of separating the star from the disk are required to probe this region.

We conclude that the SED fitting technique is sufficiently robust to reproduce the global parameters (dust mass, grain growth) of the outer disks.  We emphasize that our analysis describes reasonably well the SED for a wide range of wavelengths (3~$\mu$m--3.6~cm) with a consistent set of star, disk, and envelope parameters. Our results are also consistent with the morphology observed in the millimeter and mid-infrared images.


\begin{figure*}
 \includegraphics[viewport=0 10 500 720, width=0.45\textwidth, height=0.38\textheight, angle=-90]{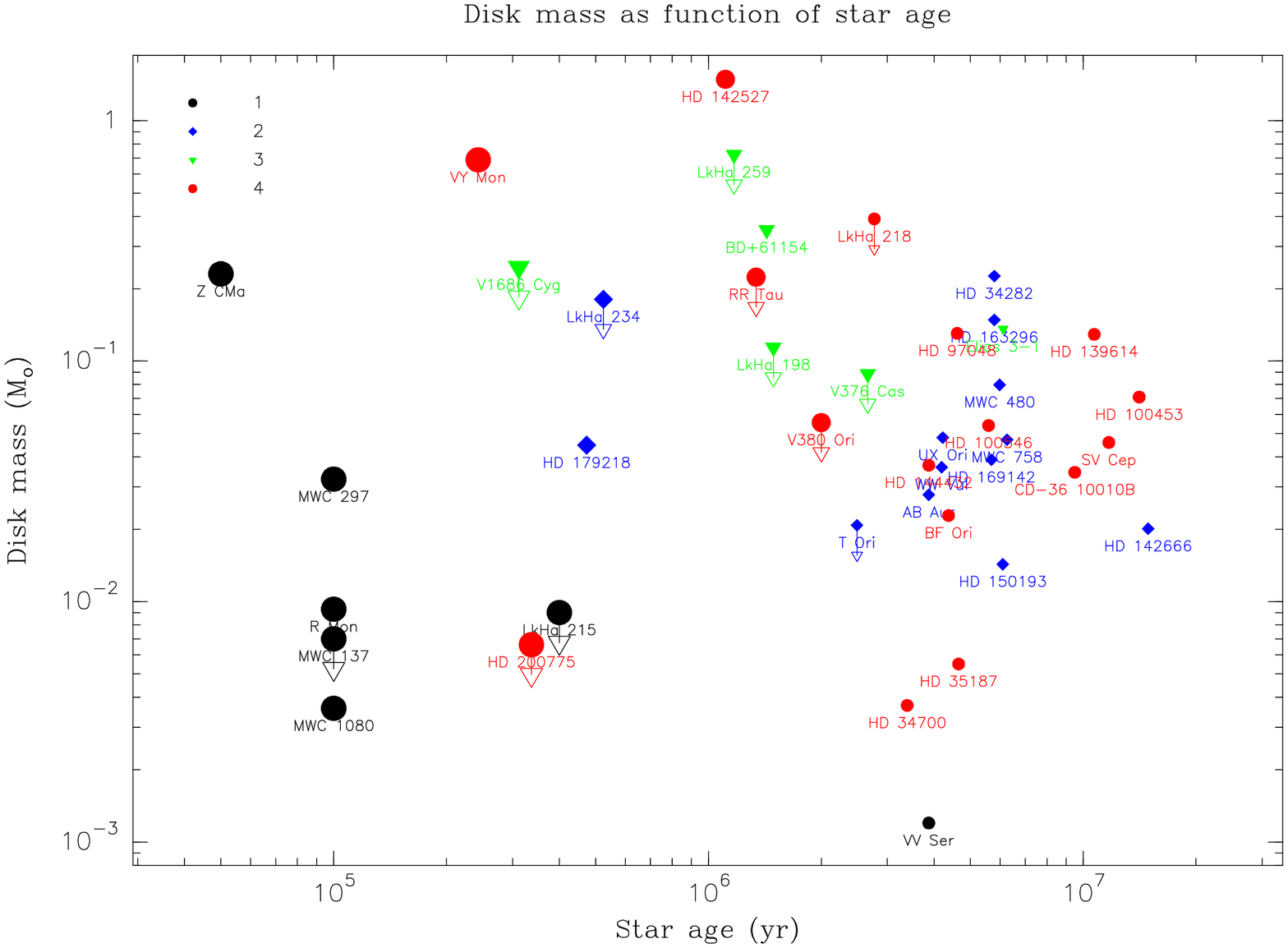}
 \includegraphics[viewport=0 10 500 720, width=0.45\textwidth, height=0.38\textheight, angle=-90]{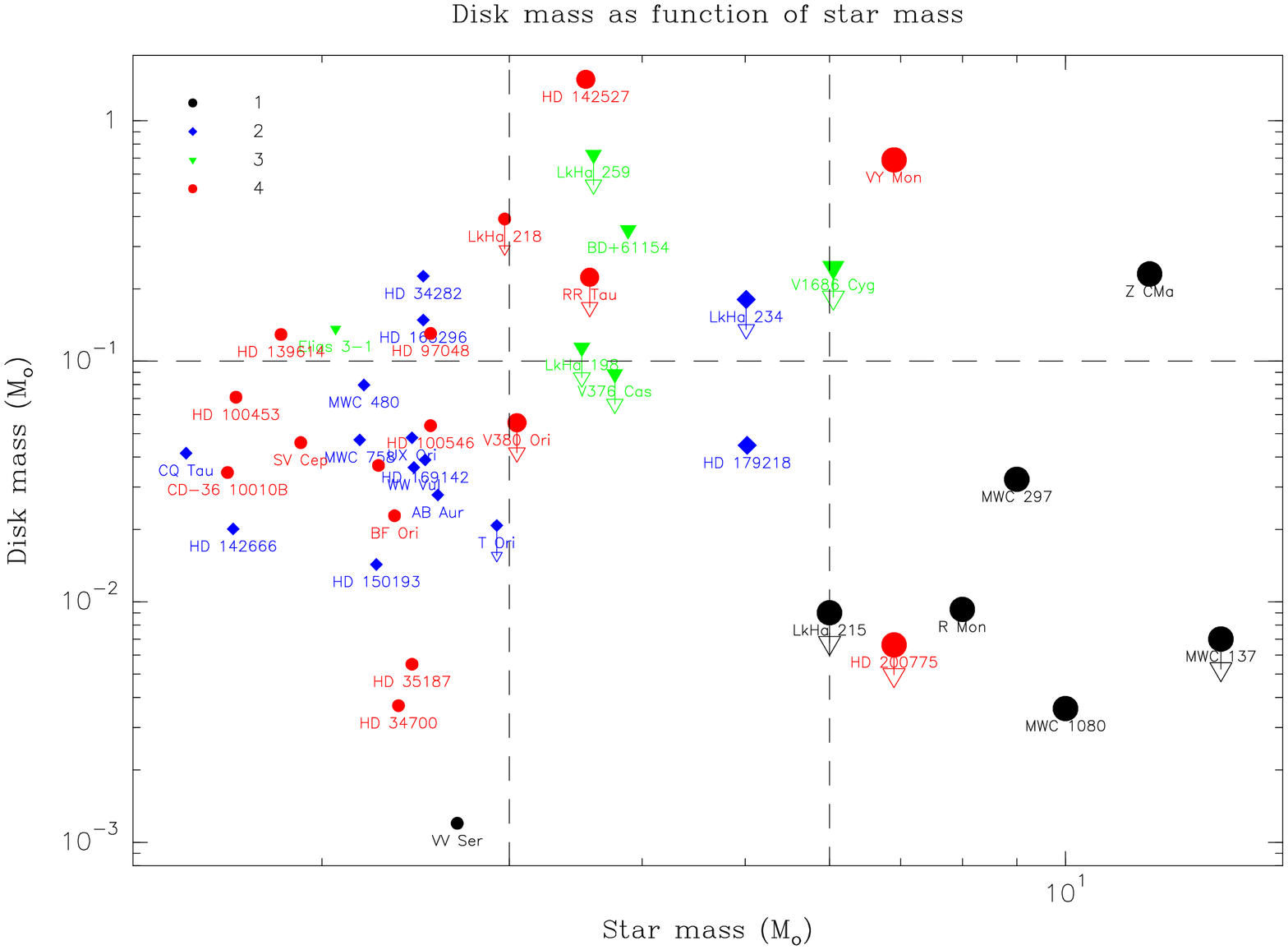}
\caption{
Compilation of disks masses (gas+dust) as function of stellar age ({\it left}) and the stellar mass ({\it right}). Our observations with black circles. In blue diamonds we represent the disk masses derived from 1.3~mm interferometric observations from other authors. With green inverted triangles we represent the disk masses derived from 2.7~mm interferometric observations. In these stars the disk mass is probably overestimated because of the possible (and not subtracted) contribution of the free-free emission. The red circles correspond to single-dish 1.3~mm observations. The disk mass in these cases could also be overestimated because of the existence of an envelope.}
\label{disksMassAge}
\label{disksMassStarMass}
\end{figure*}

\section{Discussion}

Our cm and mm interferometric observations have provided valuable insight into the structure of the circumstellar matter around HBe stars. The first result is that HBe stars are usually surrounded by a massive envelope. The mass of the envelope is always larger than the mass of the disk and dominates the emission at FIR and millimeter wavelengths. Extended emission arising in the envelope is also observed in the MIPS images at 24~$\mu$m, which shows that the envelope emission significantly contributes to the SED even at mid-IR wavelengths. The existence of a nebula was also inferred for Spitzer images towards the A0 star VV~Ser \citep{pon07}. Our observations corroborate that interferometric observations are required to derive the properties of the disks around these massive stars, and a disk+envelope model is essential when interpreting the SED.

We have different types of disks in our sample. Towards Z~CMa and R~Mon, we detected disks with sizes of r$_{out}$$>$100~AU and dust masses of 10$^{-4}$--10$^{-3}$~M$_\odot$. These values do not differ from the estimated sizes of disks around TTs and HAe stars. Both of these disks were detected in the CO J=1$\rightarrow$0 and J=2$\rightarrow$1 rotational lines (Fuente et al. 2006, Alonso-Albi et al. 2009). Z~CMa is a controversial case. 
Its accretion rate of 10$^{-3}$ M$_\odot$/yr \citep{har89} is one of the highest known around a pre-main sequence object, and Z~CMa is surrounded by a compact, massive envelope and powers an energetic outflow \citep{poe89}. 
It is probably the youngest object in our sample, and it probably contains the most massive disk. However, the dust emission could originate in a circumbinary ring and/or a disk associated with the FU Orionis companion. 

The circumstellar disk around MWC~1080 is smaller (r$_{out}$$\sim$80~AU) and less massive (5x10$^{-5}$~M$_\odot$) than those around Z~CMa and R~Mon. However, the envelope contains a significant amount of dust, which suggests that the star is still quite young. This disk has not been detected in the CO J=1$\rightarrow$0 and 2$\rightarrow$1 lines (Alonso-Albi et al. unpublished data). 

MWC~297 is a special case because its disk consists of two components. The inner disk has a radius of 28~AU, similar to the more evolved sources discussed below, while the outer ring has a radius of 200~AU, more similar to the young disks around R~Mon and Z~CMa. The disk has not been detected by its CO J=1$\rightarrow$0, J=2$\rightarrow$1 (Alonso-Albi et al. unpublished data) nor J=3$\rightarrow$2 \citep{ack08} rotational line radiation. 

Although the mass of dust in the circumstellar disk varies between one source and another (10$^{-5}$--10$^{-3}$~M$_\odot$), the 1.3mm/2.7mm spectral index is low ($<$2.2) in all sources. This 1.3mm/2.7mm spectral index indicates that grain growth has proceeded to sizes of 1~cm even in the youngest sources of our sample, R~Mon and Z~CMa. As discussed later, early grain growth has important consequences for the evolution of the dusty disk. The non-detection at mm wavelengths of the disks around MWC~137 and LkH$\alpha$~215 also implies the existence of compact ($<$20~AU) and/or presumably light disks ($<$0.001~M$_\odot$), if any, around these stars. 

\subsection{Disk masses around Herbig Ae/Be stars}
\label{statistics}
To investigate the possible dependence of the disk mass on the stellar mass and age,
we compiled a list of disks detected at mm wavelengths (1.1--2.7mm) around HAeBe stars (see Table \ref{disksTable}). It is important to calculate the disk masses in a uniform way. We calculated all disk masses, including for our own sample, by assuming optically thin emission at mm wavelengths, an average disk temperature that depends on the stellar spectral type \citep{nat00}, a dust opacity  $\kappa_{1.3mm}$=1.0~cm$^2$~g$^{-1}$, and a dust spectral index $\beta$ = 1.0. We subtracted the free-free contribution in the sources previously studied by our team \citep{fue03,fue06,alo08}. Then we calculated the disk masses by assuming that all the 1.3mm/2.7mm emission originates in the dusty disk. To estimate the uncertainty in the disk masses derived in this $``$simple$"$ way, we compared the disk masses for our sample with those previously derived by modeling the full SED in this Paper. The values agree within a factor 2-3, which suggests that our method is robust to within the statistical uncertainties. 

We also calculated the mass and age of all of these stars using the evolutionary tracks of \cite{sie00} and the values of luminosities and effective temperatures found in the literature (see Table \ref{disksTable}
). In some cases, the luminosities and effective temperatures were modified slightly (within the expected uncertainty) to fit the closest \cite{sie00} evolutionary track and the spectral type of each star. Stars with masses higher than 7~M$_\odot$ do not undergo a pre-main sequence phase and their ages cannot be estimated using the HR diagram. In these cases, we evaluated the age on the basis of different arguments. For Z~CMa, we assumed an age of 0.05 Myr consistent with the time required to form a star of 16~M$_\odot$ with a mean accretion rate of 3x10$^{-4}$~M$_\odot$~yr$^{-1}$. This star could have been even younger if the accretion rate had remained close to 10$^{-3}$~M$_\odot$/yr throughtout the formation process. MWC~137, MWC~297, MWC~1080, and R~Mon are very likely young main-sequence objects. For these objects, we assumed that the age was lower than 1~Myr based on the post-main sequence tracks of \cite{sch92}. For clarity and to differentiate them from other sources in which we have been able to estimate stellar ages, we plot them in the 0.1~Myr slot in the disk mass versus stellar age diagram.

In Fig. \ref{disksMassStarMass} (right panel), we present the total disk masses (gas+dust) as a function of the stellar mass assuming a gas/dust ratio of 100.  We can easily distinguish three regions in the plot. Firstly, for stars with masses of $<$3~M$_\odot$, there is a cloud of points around a mean value of $\sim$0.04~M$_\odot$. However, there are a few stars with light disks ($<$0.01~$M_\odot$). Secondly, in stars with masses of 3-7 M$_\odot$, there is a group of massive disks ($>$0.1~M$_\odot$). Most of these disks masses were derived from 2.7mm observations and were probably overestimated because of the unsubtracted free-free emission. In the case of HD~142527 and VY~Mon, the disk masses were derived from single-dish observations and we cannot discard the contribution of an associated nebula. 
Thus, these massive disks requires further confirmation. Thirdly, in more massive stars ($>$7~M$_\odot$), mainly our entire sample of HBe stars, the disk masses decrease by a factor 5-10 and become $<$0.01~M$_\odot$. The only exception is Z~CMa, a particularly young object as discussed above. Since it is a binary containing a 16~M$_\odot$ and 3~$M_\odot$ stars, the circumstellar disk could be associated with the lower mass component and/or a circumbinary disk. Another interesting case is MWC~297. If we consider only the inner disk, it is located in the region of the $``$light disks$"$ ($<$0.01~M$_\odot$) related to massive stars. If we consider the sum of the two disk components, the disk mass is similar to those of the disks associated with HAe stars.

In Fig. \ref{disksMassAge} (left panel), we show the disk masses as a function of the stellar age. We can also distinguish the three regions described above. The cloud of points of stars with masses $<$3~M$_\odot$ and disk masses $\sim$0.04~M$_\odot$ corresponds to stars of ages between a few Myr and 10 Myr. The most massive disks corresponds to stars of masses between 3--7~M$_\odot$ and ages of around 1~Myr. The circumstellar disks around stars with masses of $>$7~M$_\odot$ exhibit a large mass dispersion, ranging from $>$0.1~M$_\odot$ to $<$0.01~M$_\odot$, i.e., two orders of magnitude, although mainly at the lower end of this range.

To attempt to understand this behavior, we recall that we only consider pre-main sequence (visible) objects. Thus, there is a lower limit to the age of the star that corresponds to the birthline and depends on the stellar mass. The upper limit to the stellar ages for our disk detections is given by the disk lifetime which probably also depends on stellar mass. For Herbig Ae stars with masses between 1--3~M$_\odot$, this period (from the birthline until the disk disperses) corresponds to stellar ages of between a few Myr and 10 Myr. During this phase the disk mass does not vary significantly with stellar age. For stars with masses in the range 3--7~M$_\odot$, this phase lasts from 10$^{5}$~yr to $\sim$2~Myr.  For the most massive stars, $>$7~M$_\odot$,  once the star is visible the disk dispersal is so rapid that instead of having a cloud of points around a mean value, we have a large dispersion of the disk masses (2 orders of magnitude), and most of objects in the lower end (M$\leq$0.01~M$_\odot$).
This indicates that, although massive disks are found towards young, intermediate-mass protostars \citep{ces05,sch06} and possibly Z~CMa, after 1~Myr massive stars have already dispersed most of the circumstellar material and are surrounded by disks of masses lower than 0.01~M$_\odot$.

\begin{table}[h]
\caption{Dissipation timescale.} 
\begin{tabular} 
{@{\extracolsep{\fill}}llll}
\hline\hline\noalign{\smallskip}
Source & $R_g$$^1$ & Mass loss rate & Life time$^2$ \\
       & (AU) & (M$_\odot$/yr) & (yr) \\
\hline\noalign{\smallskip}
MWC~137 & 124.6 & 8.9x10$^{-6}$ & 1.1x10$^5$   \\
LKH$\alpha$~215 & 53.4 & 3.4x10$^{-6}$ & 2.9x10$^5$\\
R~Mon & 71.2 & 4.1x10$^{-6}$ & 2.4x10$^5$ \\
Z~CMa & 106.8 & 6.1x10$^{-6}$ &  1.6x10$^5$ \\
MWC~297 & 80 & 6.0x10$^{-6}$ & 1.7x10$^5$ \\
MWC~1080 & 89 & 4.6x10$^{-6}$ &  2.2x10$^5$ \\
\noalign{\smallskip}\hline\hline
\end{tabular}

\noindent
$^1$ Gravitational radius calculated following expression (1) by \cite{ale07} 

\noindent
$^2$ Time required to disperse a 1~M$_\odot$ disk.
\label{disipation}
\end{table}

\subsection{Disk photo-evaporation in massive stars}
Different dynamical processes contribute to the dispersal of gas in circumstellar disks: viscosity, winds and jets, dynamical encounters with nearby stars, and photoevaporation by the central star and external illumination. The relative importance of these processes was discussed by \cite{hol00}, who concluded that the dominant processes affecting disk evolution are viscosity in the inner regions ($<$10~AU) and photoevaporation at larger radii. Photoevaporation affects firstly the outer region of the disk, where the gravitational field is lower and the thermal heating can disrupt the disk (see e.g. the review by Alexander, 2007). We can define a characteristic radius R$_g$ beyond which material evaporates and a gap opens in the disk. As soon as the gap opens, the direct photoevaporation disperses the gaseous disk from the inside to the outside in a few 10$^5$ yr.

While the gas evolution is outlined above, the dust evolution is more difficult to predict and depends on the maximum grain size \citep{gar07}. Particles are coupled to the gas by means of the frictional drag that depends on grain size. If the particles are far smaller than the mean free path of the gas molecules ($\lambda_{mfp}$), gas and grains are coupled. Assuming typical gas densities, the grains with sizes smaller than 1~$\mu$m remain coupled to the gas. Thus, if grain growth is ignored, the evolution of the solid particles relates to that of the gas and the dusty disk is rapidly dispersed once the gap is opened. However, if grain growth is taken into account the evolution of the dusty disk is different. Although small particles are blown away with the gas, the large particles remain forming a compact, dusty disk. \cite{alearm07} showed that there is a radius beyond which all solids are entrained in the gas. This radius is identified with the region where the maximum grain size is between 1--10~cm.

We propose that the presence of compact, dusty disks around most HBe stars is the consequence of rapid disk photoevaporation in these massive stars. Since the ionizing flux in HBe stars is higher than in the cooler HAe and TT stars, the timescale for disk photoevaporation is shorter.
In Table \ref{disipation} we show the characteristic gap radius and the photoevaporation rate for the stars in our sample calculated following \cite{ale07}. The mass-loss rate is about an order of magnitude higher in Be stars than in Ae stars.  While VV~Ser as an A0 star needs about 9~Myr to dissipate 1~M$_\odot$ disk mass, HBe stars disperse their gaseous disk on a few 10$^5$ yr. This time is so short that the star could dissipate the outer part of the disk before becoming visible. As soon as the outer disk is dispersed, the entire gaseous disk is dispersed on a short timescale. This is consistent with the lack of CO detection in the small disks, MWC~1080 and MWC~297, of our sample. One important result of our mm-wave analysis is that grain growth begins early in disk evolution, and grains have grown to sizes of $\sim$~1~cm even before the star becomes visible. Thus, once the gaseous disk is dispersed, a compact, dusty disk formed by large grains remains. An interesting case is MWC~297 in which we find a small, circumstellar disk (r$_{out}$=29~AU) and a remnant, outer ring with r$_{in}$=200~AU and r$_{out}$=300~AU. This second ring could be the consequence of the decrease in gas density outwards from the star centre. The grains entrained in the retreating gas are left behind when the gas density is too low for the drag force to push them away. MWC~297 could be an object in-between the young disks associated with R~Mon and Z~CMa and the compact, dusty disk surrounding MWC~1080 and presumably, MWC~137 and LkH$\alpha$~215.

Additional support for the rapid photoevaporation of disks in HBe stars is found by \cite{ber08} on the basis of the PAH and small grains (VSG) emission in the surroundings of HBe stars. They suggested that the PAH/VSG emission from disks around B stars is absent. The observed emission emanates from the PDRs formed in the remnant of the parental cloud. They interpret this lack of emission as a sign that PAHs/VSGs have been photodissociated by hard UV photons in the disks. Since PAHs represent most of the UV optical depth, we conclude that this is also in good agreement with a scenario of photoevaporation in the disk around massive stars.


\section{Conclusions}

We briefly summarize the main conclusions of our study:

\begin{itemize}
\item We have carried out a search for circumstellar disks around HBe stars using the NRAO 
Very Large Array (VLA) and IRAM Plateau de Bure (PdB) interferometers. Interferometric observations
are necessary to separate the disk from the surrounding envelope, although the disks themselves usually 
remain unresolved. Our goal is to investigate 
the properties of the circumstellar disks around intermediate mass stars to determine eventually their occurrence, lifetime, and evolution. Thus far, we have observed 6 objects with 4 successful detections.

\item
Our first result is that the disk mass is usually only a small percentage (less than 10\%) of the mass of the entire envelope. There are significant variations in the disk mass from source to source. Two disks of our sample, R~Mon and Z~CMa, have similar sizes to those found in TT and HAe stars. Z~CMa is a FU Orionis object and cannot be directly compared with the others. The mass of the other large disk, R~Mon, is a factor of 5 below that of an average HAe star. The disk around MWC~1080 is smaller (r$_ {out}$$<$100~AU) and less massive. Around MWC~297, we find an inner compact disk (r$_{out}$$\sim$~29~AU) and an outer ring at a distance of 200~AU. The 1.3mm/2.7mm spectral index indicates that grain growth has proceeded in all disks in our sample.

\item
A comparison between our data and previous results in TT and HAe stars show that the masses of the circumstellar disks around HBe stars are on average 5-10 times lower than those around Herbig Ae stars. We propose that disk photoevaporation is responsible for this behavior. In HBe stars, the UV radiation disperses the gas in the outer disk on a timescale of a few 10$^5$ yr. Once the outer part of the disk dissapears, the entire gaseous disk is photoevaporated on a short timescale ($\sim$10$^5$ yr), and only a dusty disk composed of large grains remains.
\end{itemize}

\begin{acknowledgement}
We thank the IRAM staff in Grenoble for their help and support during the observations and data reduction. 
This work was partly supported by INAF PRIN 2006 grant $``$From disks to planetary systems$"$. We thank
the referee for his/her fruitful comments.
\end{acknowledgement}


\clearpage
\newpage


\begin{appendix} 

 \section{}

\begin{table*}[h]
\caption{Table of fluxes for MWC 137.} 
\begin{tabular*}{1.0\textwidth}
{@{\extracolsep{\fill}}lllll}
\hline\hline\noalign{\smallskip}
Wavelength ($\mu m$) & Flux (Jy) & Flux error (Jy) & Beam size ($''$) & Reference / Flux type \\
\hline\noalign{\smallskip}
0.440 & 0.019 & 0.001 & 0.8 & TYCHO-2 BAND$\_$B \\
0.550 & 0.081 & 0.001 & 0.8 & TYCHO-2 BAND$\_$V \\
0.700 & 0.017 & 0.003 & 0.9 & USNO-B BAND$\_$R$\_$JOHNSON$\_$MORGAN \\
0.900 & 0.012 & 0.002 & 0.9 & USNO-B BAND$\_$I$\_$JOHNSON$\_$MORGAN \\
1.235 & 0.499 & 0.010 & 2 & BAND$\_$J$\_$2MASS \\
1.620 & 0.428 & 0.001 & 1 & Morel 1978 BAND$\_$H$\_$JOHNSON$\_$MORGAN \\
1.662 & 0.749 & 0.017 & 2 & BAND$\_$H$\_$2MASS \\
2.159 & 1.500 & 0.030 & 2 & BAND$\_$Ks$\_$2MASS \\
2.200 & 1.719 & 0.001 & 1 & Morel 1978 BAND$\_$K$\_$JOHNSON$\_$MORGAN \\
3.400 & 3 & 0 & 1 & Morel 1978 BAND$\_$L$\_$JOHNSON$\_$MORGAN \\
8.280 & 4.308 & 0.01 & 18.3 & BAND$\_$A$\_$MSX6C \\
12.000 & 10.9 & 0.1 & 30 & Hillenbrand 1992 (IRAS) \\
12.130 & 4.371 & 0.010 & 18.3 & BAND$\_$C$\_$MSX6C \\
14.650 & 2.713 & 0.010 & 18.3 & BAND$\_$D$\_$MSX6C \\
21.340 & 4.834 & 0.010 & 18.3 & BAND$\_$E$\_$MSX6C \\
25.000 & 25.7 & 0.1 & 30 & Hillenbrand 1992 (IRAS) \\
60.000 & 245.0 & 1.0 & 60 & Hillenbrand 1992 (IRAS) \\
100 & 519.0 & 1.0 & 120 & Hillenbrand 1992 (IRAS) \\
350 & 4 & 1 & 18.5 & Mannings 1994 \\
450 & 1.66 & 0.25 & 18.3 & Mannings 1994 \\
750 & 0.300 & 0.060 & 18.0 & Mannings 1994 \\
800 & 0.370 & 0.050 & 17.0 & Mannings 1994 \\
1100 & 0.095 & 0.014 & 18.8 & Mannings 1994 \\
1300 & 0.090 & 0.013 & 19.8 & Mannings 1994 \\
1400 & 0.0071 & 6$x10^{-4}$ & 3.05x1.78 & Our data \\
2700 & 0.0041 & 2.0$x10^{-4}$ & 6.2x3.63 & Our data \\
7000 & 2.35 $10^{-3}$ & 1.3$x10^{-4}$ & 2.21x1.66 & Our data \\
13000 & 0.0024 & 5$x10^{-4}$ & 5.18x3.19 & Our data \\
36000 & 5.7$x10^{-4}$ & 5$x10^{-5}$ & 0.3 & Skinner 1993 \\
60000 & $<$6.0$x10^{-4}$ & - & 1.7 & Skinner 1993 \\
\noalign{\smallskip}\hline\hline
\end{tabular*}
\label{mwc137}
\end{table*}

\begin{table*}[h]
\caption{Extinction details for MWC 137.} 
\begin{tabular*}{1.0\textwidth}
{@{\extracolsep{\fill}}lllll}
\hline\hline\noalign{\smallskip}
Wavelength & Extinction & Extinction from envelope model & Observed flux & Dereddened flux \\
($\mu m$) & (mag) & (mag) & (mJy) & (mJy) \\
\hline\noalign{\smallskip}
0.44 & 7.80 & 1.05 & 19.2 & 25340.3 \\
0.55 & 5.88 & 0.79 & 81.1 & 18294.1 \\
0.7 & 4.42 & 0.59 & 17.3 & 1013.9 \\
0.9 & 2.82 & 0.38 & 11.6 & 156.2 \\
1.235 & 1.69 & 0.23 & 498.5 & 2372.9 \\
1.62 & 1.09 & 0.15 & 428.1 & 1173.0 \\
1.662 & 1.05 & 0.14 & 748.7 & 1969.6 \\
2.159 & 0.69 & 0.09 & 1495.3 & 2821.1 \\
2.2 & 0.67 & 0.09 & 1719.3 & 3182.7 \\
\noalign{\smallskip}\hline\hline
\end{tabular*}
\label{mwc137_ext}
\end{table*}

\clearpage
\newpage

\begin{table*}[h]
\caption{Table of fluxes for LKH$\alpha$~215.} 
\begin{tabular*}{1.0\textwidth}
{@{\extracolsep{\fill}}lllll}
\hline\hline\noalign{\smallskip}
Wavelength ($\mu m$) & Flux (Jy) & Flux error (Jy) & Beam size ($''$) & Reference / Flux type \\
\hline\noalign{\smallskip}
0.360 & 0.043 & 0.002 & 1 & Reed 2005 BAND$\_$U$\_$JOHNSON$\_$MORGAN \\
0.440 & 0.110 & 0.001 & 1 & AC2000.2 BAND$\_$B$\_$JOHNSON$\_$MORGAN \\
0.550 & 0.166 & 0.007 & 1 & Hipparcos BAND$\_$V$\_$JOHNSON$\_$MORGAN \\
0.550 & 0.219 & 0.009 & 1 & Corporon 1999 BAND$\_$V$\_$JOHNSON$\_$MORGAN \\
0.700 & 0.208 & 0.008 & 1 & Zacharias 2005 BAND$\_$R$\_$JOHNSON$\_$MORGAN \\
1.235 & 0.580 & 0.012 & 2 & BAND$\_$J$\_$2MASS \\
1.620 & 0.328 & 0.001 & 1 & Morel 1978 BAND$\_$H$\_$JOHNSON$\_$MORGAN \\
1.662 & 0.762 & 0.020 & 2 & BAND$\_$H$\_$2MASS \\
2.159 & 1.024 & 0.020 & 2 & BAND$\_$Ks$\_$2MASS \\
2.200 & 1.036 & 0.001 & 1 & Morel 1978 BAND$\_$K$\_$JOHNSON$\_$MORGAN \\
3.400 & 1.280 & 0.001 & 1 & Morel 1978 BAND$\_$L$\_$JOHNSON$\_$MORGAN \\
8.280 & 3.261 & 0.01 & 18.3 & BAND$\_$A$\_$MSX6C \\
12.000 & 6.3 & 0.1 & 30 & Hillenbrand 1992 (IRAS) \\
12.130 & 4.2301 & 0.01 & 18.3 & BAND$\_$C$\_$MSX6C \\
14.650 & 1.84 & 0.01 & 18.3 & BAND$\_$D$\_$MSX6C \\
21.340 & 2.15 & 0.01 & 18.3 & BAND$\_$E$\_$MSX6C \\
25.000 & 8.2 & 0.1 & 30 & Hillenbrand 1992 (IRAS) \\
60.000 & 86.9 & 0.1 & 60 & Hillenbrand 1992 (IRAS) \\
100 & 315.0 & 1.0 & 120 & Hillenbrand 1992 (IRAS) \\
450 & 1.23 & 0.22 & 18.3 & Mannings 1994 \\
800 & 0.142 & 0.020 & 17.0 & Mannings 1994 \\
1100 & 0.058 & 0.014 & 18.8 & Mannings 1994 \\
1400 & $<$0.0010 & - & 3.32x1.84 & Our data \\
2700 & $<$3.0$x10^{-4}$ & - & 6.77x3.68 & Our data \\
\noalign{\smallskip}\hline\hline
\end{tabular*}
\label{lkha215}
\end{table*}

\begin{table*}[h]
\caption{Extinction details for LKH$\alpha$~215.} 
\begin{tabular*}{1.0\textwidth}
{@{\extracolsep{\fill}}lllll}
\hline\hline\noalign{\smallskip}
Wavelength & Extinction & Extinction from envelope model & Observed flux & Dereddened flux \\
($\mu m$) & (mag) & (mag) & (mJy) & (mJy) \\
\hline\noalign{\smallskip}
0.36 & 5.37 & 0.78 & 42.7 & 6009.6 \\
0.44 & 4.54 & 0.66 & 109.5 & 7148.9 \\
0.55 & 3.42 & 0.50 & 166.3 & 3886.4 \\
0.55 & 3.42 & 0.50 & 219.2 & 5123.3 \\
0.70 & 2.57 & 0.37 & 208.2 & 2220.5 \\
1.235 & 0.99 & 0.14 & 579.8 & 1436.7 \\
1.62 & 0.64 & 0.09 & 327.8 & 589.0 \\
1.662 & 0.61 & 0.09 & 761.9 & 1337.3 \\
2.159 & 0.40 & 0.06 & 1024.1 & 1481.4 \\
2.20 & 0.39 & 0.06 & 1036.0 & 1482.1 \\

\noalign{\smallskip}\hline\hline
\end{tabular*}
\label{lkha215_ext}
\end{table*}

\clearpage
\newpage

\begin{table*}[h]
\caption{Table of fluxes for R Mon.} 
\begin{tabular*}{1.0\textwidth}
{@{\extracolsep{\fill}}lllll}
\hline\hline\noalign{\smallskip}
Wavelength ($\mu m$) & Flux (Jy) & Flux error (Jy) & Beam size ($''$) & Reference / Flux type \\
\hline\noalign{\smallskip}
1.235 & 0.213 & 0.005 & 2 & BAND$\_$J$\_$2MASS \\
1.256 & 0.027 & 0.0010 & 6 & Close 1997 \\
1.633 & 0.25 & 0.01 & 6 & Close 1997 \\
1.662 & 0.630 & 0.015 & 2 & BAND$\_$H$\_$2MASS \\
2.159 & 1.860 & 0.040 & 2 & BAND$\_$Ks$\_$2MASS \\
2.210 & 0.99 & 0.01 & 6 & Close 1997 \\
3.163 & 11.17 & 0.01 & 6 & Close 1997 \\
3.600 & 18 & 0 & 12 & Audard 2007 \\
3.930 & 19.69 & 0.01 & 6 & Close 1997 \\
4.290 & 21.05 & 0.01 & 18.3 & BAND$\_$B1$\_$MSX6C \\
4.350 & 29.16 & 0.01 & 18.3 & BAND$\_$B2$\_$MSX6C \\
4.674 & 23.85 & 0.01 & 6 & Close 1997 \\
5.800 & 30.41 & 0.4 & 12 & Audard 2007 \\
8.000 & 31 & 0 & 12 & Audard 2007 \\
8.280 & 32.09 & 0.01 & 18.3 & BAND$\_$A$\_$MSX6C \\
11.700 & 46.59 & 0.01 & 6 & Close 1997 \\
12.000 & 58.3 & 0.1 & 30 & Hillenbrand 1992 (IRAS) \\
12.130 & 47.18 & 0.01 & 18.3 & BAND$\_$C$\_$MSX6C \\
14.650 & 58.53 & 0.01 & 18.3 & BAND$\_$D$\_$MSX6C \\
20.800 & 103.1 & 0.1 & 6 & Close 1997 \\
21.340 & 102.2 & 0.01 & 18.3 & BAND$\_$E$\_$MSX6C \\
25.000 & 139.0 & 1.0 & 30 & Hillenbrand 1992 (IRAS) \\
52.560 & 99 & 4 & 30 & Spitzer .tbl file \\
69.660 & 81 & 3 & 30 & Spitzer .tbl file \\
93.600 & 62.3 & 2.0 & 30 & Spitzer .tbl file \\
350 & 2 & 0.300 & 18.5 & Mannings 1994 \\
450 & 1.56 & 0.25 & 18.3 & Mannings 1994 \\
800 & 0.220 & 0.040 & 17.0 & Mannings 1994 \\
1100 & 0.077 & 0.018 & 18.8 & Mannings 1994 \\
1200 & 0.049 & $10^{-4}$ & 11 & Our data - 30m bolometer map \\
1300 & 0.079 & 0.019 & 19.8 & Mannings 1994 \\
1400 & 0.0118 & 0.0020 & 0.72x0.33 & Our data \\
2700 & 0.0041 & 5.0$x10^{-4}$ & 1.28x0.83 & Our data \\
7000 & 1.26$x10^{-3}$ & 1.3$x10^{-4}$ & 2.21x1.66 & Our data \\
13000 & 7.8$x10^{-4}$ & 1.0$x10^{-4}$ & 5.18x3.19 & Our data \\
20000 & 5.9$x10^{-4}$ & 3.0$x10^{-4}$ & 1.4 & Skinner 1993 \\
36000 & 4.0$x10^{-4}$ & 1.0$x10^{-4}$ & 0.3 & Skinner 1993 \\
60000 & 2.9$x10^{-4}$ & 1.0$x10^{-4}$ & 4.3 & Skinner 1993 \\
\noalign{\smallskip}\hline\hline
\end{tabular*}
\label{rmon}
\end{table*}

\begin{table*}[h]
\caption{Extinction details for R Mon.} 
\begin{tabular*}{1.0\textwidth}
{@{\extracolsep{\fill}}lllll}
\hline\hline\noalign{\smallskip}
Wavelength & Extinction & Extinction from envelope model & Observed flux & Dereddened flux \\
($\mu m$) & (mag) & (mag) & (mJy) & (mJy) \\
\hline\noalign{\smallskip}
1.235 & 0.66 & 1.36 & 212.9 & 390.8 \\
1.256 & 0.64 & 1.33 & 27.0 & 48.8 \\
1.633 & 0.42 & 0.87 & 250.0 & 368.3 \\
1.662 & 0.41 & 0.85 & 629.7 & 917.7 \\
2.159 & 0.27 & 0.55 & 1863.5 & 2386.2 \\
2.210 & 0.26 & 0.53 & 990.0 & 1256.2 \\
3.163 & 0.15 & 0.30 & 11170.0 & 12767.7 \\
\noalign{\smallskip}\hline\hline
\end{tabular*}
\label{rmon_ext}
\end{table*}

\clearpage
\newpage

\begin{table*}[h]
\caption{Table of fluxes for Z CMa.} 
\begin{tabular*}{1.0\textwidth}
{@{\extracolsep{\fill}}lllll}
\hline\hline\noalign{\smallskip}
Wavelength ($\mu m$) & Flux (Jy) & Flux error (Jy) & Beam size ($''$) & Reference / Flux type \\
\hline\noalign{\smallskip}
0.440 & 0.205 & 0.001 & 0.8 & TYCHO-2 BAND$\_$B \\
0.550 & 0.474 & 0.001 & 0.8 & TYCHO-2 BAND$\_$V \\
0.700 & 0.7 & 0.3 & 0.9 & USNO-B BAND$\_$R$\_$JOHNSON$\_$MORGAN \\
0.900 & 0.9 & 0.4 & 0.9 & USNO-B BAND$\_$I$\_$JOHNSON$\_$MORGAN \\
1.235 & 4 & 0 & 2 & BAND$\_$J$\_$2MASS \\
1.620 & 6 & 0 & 1 & Morel 1978 BAND$\_$H$\_$JOHNSON$\_$MORGAN \\
1.662 & 8 & 0 & 2 & BAND$\_$H$\_$2MASS \\
2.159 & 21 & 3 & 2 & BAND$\_$Ks$\_$2MASS \\
2.200 & 25 & 0 & 1 & Morel 1978 BAND$\_$K$\_$JOHNSON$\_$MORGAN \\
3.400 & 52 & 0 & 1 & Morel 1978 BAND$\_$L$\_$JOHNSON$\_$MORGAN \\
3.800 & 54.0 & 4.0 & $<$0.1 & Koresko 1991 \\
4.290 & 71.78 & 0.01 & 18.3 & BAND$\_$B1$\_$MSX6C \\
4.350 & 72.12 & 0.01 & 18.3 & BAND$\_$B2$\_$MSX6C \\
4.800 & 69.0 & 6.0 & $<$0.1 & Koresko 1991 \\
8.280 & 107.7 & 0.1 & 18.3 & BAND$\_$A$\_$MSX6C \\
10.000 & 109.0 & 12.0 & $<$0.1 & Koresko 1991 \\
12.000 & 126.6 & 0.1 & 30 & Oudmaijer 1992 (IRAS) \\
12.130 & 140.2 & 0.1 & 18.3 & BAND$\_$C$\_$MSX6C \\
14.650 & 159.1 & 0.1 & 18.3 & BAND$\_$D$\_$MSX6C \\
20.000 & 183.0 & 5.0 & 10 & ISO \\
21.340 & 202.8 & 0.1 & 18.3 & BAND$\_$E$\_$MSX6C \\
25.000 & 221.3 & 0.1 & 30 & Oudmaijer 1992 (IRAS) \\
40.000 & 265.0 & 5.0 & 20 & ISO \\
50.000 & 340.0 & 10.0 & 30 & Elia 2004 (ISO, chart estimate) \\
60.000 & 322.0 & 0.1 & 60 & Oudmaijer 1992 (IRAS) \\
60.000 & 365.0 & 10.0 & 40 & Elia 2004 (ISO, chart estimate) \\
70.000 & 390.0 & 10.0 & 40 & Elia 2004 (ISO, chart estimate) \\
80.000 & 410.0 & 10.0 & 60 & Elia 2004 (ISO, chart estimate) \\
90.000 & 420.0 & 10.0 & 80 & Elia 2004 (ISO, chart estimate) \\
100 & 354.0 & 0.1 & 120 & Oudmaijer 1992 (IRAS) \\
100 & 410.0 & 10.0 & 80 & Elia 2004 (ISO, chart estimate) \\
120 & 390.0 & 10.0 & 80 & Elia 2004 (ISO, chart estimate) \\
140 & 380.0 & 10.0 & 80 & Elia 2004 (ISO, chart estimate) \\
160 & 310.0 & 10.0 & 80 & Elia 2004 (ISO, chart estimate) \\
180 & 270.0 & 10.0 & 100 & Elia 2004 (ISO, chart estimate) \\
350 & 29 & 1 & 19.0 & Dent 1998 \\
450 & 13.84 & 0.2 & 17.5 & Dent 1998 \\
450 & 11.41 & 0.1 & 8.0 & Sandell 2001 \\
800 & 1.964 & 0.013 & 16.0 & Dent 1998 \\
850 & 1.5 & 0.1 & 15.0 & Sandell 2001 \\
1100 & 0.710 & 0.030 & 18.7 & Dent 1998 \\
1300 & 0.83 & 0.01 & 21.0 & Sandell 2001 \\
1400 & 0.0260 & 0.0010 & 2.19x0.85 & Our data \\
2700 & 0.0085 & 7.0$x10^{-4}$ & 3.05x1.41 & Our data \\
6917 & 0.0021 & 1.0$x10^{-4}$ & 1.44x1.43 & Our data \\
13350 & 0.0021 & 1.0$x10^{-4}$ & 3.53x2.71 & Our data \\
36000 & 0.0022 & 3.0$x10^{-4}$ & 9.18x6.56 & Our data \\
\noalign{\smallskip}\hline\hline
\end{tabular*}
\label{zcma}
\end{table*}

\begin{table*}[h]
\caption{Extinction details for Z CMa.} 
\begin{tabular*}{1.0\textwidth}
{@{\extracolsep{\fill}}lllll}
\hline\hline\noalign{\smallskip}
Wavelength & Extinction & Extinction from envelope model & Observed flux & Dereddened flux \\
($\mu m$) & (mag) & (mag) & (mJy) & (mJy) \\
\hline\noalign{\smallskip}
0.44 & 5.09 & 0.0 & 204.6 & 22253.1 \\
0.55 & 3.84 & 0.0 & 474.0 & 16279.5 \\
0.7 & 2.88 & 0.0 & 689.6 & 9818.2 \\
0.9 & 1.84 & 0.0 & 890.4 & 4850.9 \\
1.235 & 1.11 & 0.0 & 3848.5 & 10654.0 \\
1.62 & 0.71 & 0.0 & 6075.0 & 11728.6 \\
1.662 & 0.69 & 0.0 & 8392.7 & 15778.0 \\
2.159 & 0.45 & 0.0 & 20774.5 & 31437.6 \\
2.2 & 0.44 & 0.0 & 24850.8 & 37144.3 \\
\noalign{\smallskip}\hline\hline
\end{tabular*}
\label{zcma_ext}
\end{table*}

\clearpage
\newpage

\begin{table*}[h]
\caption{Table of fluxes for MWC 297.} 
\begin{tabular*}{1.0\textwidth}
{@{\extracolsep{\fill}}lllll}
\hline\hline\noalign{\smallskip}
Wavelength ($\mu m$) & Flux (Jy) & Flux error (Jy) & Beam size ($''$) & Reference / Flux type \\
\hline\noalign{\smallskip}
0.440 & 0.012 & 0.002 & 0.9 & USNO-B BAND$\_$B$\_$JOHNSON$\_$MORGAN \\
0.550 & 0.046 & 0.002 & $<$1 & Eisner 2004 BAND$\_$V$\_$JOHNSON$\_$MORGAN \\
0.700 & 0.300 & 0.060 & 0.9 & USNO-B BAND$\_$R$\_$JOHNSON$\_$MORGAN \\
0.791 & 0.609 & 0.005 & 3 & BAND$\_$I$\_$DENIS \\
0.900 & 1.040 & 0.210 & 0.9 & USNO-B BAND$\_$I$\_$JOHNSON$\_$MORGAN \\
1.228 & 8 & 0 & 3 & BAND$\_$J$\_$DENIS \\
1.235 & 6 & 0 & 2 & BAND$\_$J$\_$2MASS \\
1.620 & 6 & 0 & 1 & Morel 1978 BAND$\_$H$\_$JOHNSON$\_$MORGAN \\
1.662 & 18 & 2 & 2 & BAND$\_$H$\_$2MASS \\
2.159 & 41 & 4 & 2 & BAND$\_$Ks$\_$2MASS \\
2.200 & 37 & 0 & 1 & Morel 1978 BAND$\_$K$\_$JOHNSON$\_$MORGAN \\
3.400 & 78 & 0 & 1 & Morel 1978 BAND$\_$L$\_$JOHNSON$\_$MORGAN \\
4.290 & 97.89 & 0.01 & 18.3 & BAND$\_$B1$\_$MSX6C \\
4.350 & 90.04 & 0.01 & 18.3 & BAND$\_$B2$\_$MSX6C \\
8.280 & 141.2 & 0.01 & 18.3 & BAND$\_$A$\_$MSX6C \\
12.000 & 159.0 & 1.0 & 30 & Hillenbrand 1992 (IRAS) \\
12.130 & 124.7 & 0.01 & 18.3 & BAND$\_$C$\_$MSX6C \\
14.650 & 104.8 & 0.01 & 18.3 & BAND$\_$D$\_$MSX6C \\
21.340 & 114.2 & 0.01 & 18.3 & BAND$\_$E$\_$MSX6C \\
25.000 & 224.0 & 1.0 & 30 & Hillenbrand 1992 (IRAS) \\
60.000 & 914.0 & 1.0 & 60 & Hillenbrand 1992 (IRAS) \\
100 & 1800.0 & 10.0 & 120 & Hillenbrand 1992 (IRAS) \\
350 & 4 & 0 & 18.5 & Mannings 1994 \\
450 & 2.46 & 0.13 & 18.3 & Mannings 1994 \\
600 & 0.970 & 0.110 & 17.5 & Mannings 1994 \\
750 & 0.840 & 0.060 & 18.0 & Mannings 1994 \\
800 & 0.699 & 0.013 & 17.0 & Mannings 1994 \\
850 & 0.681 & 0.022 & 18.0 & Mannings 1994 \\
1100 & 0.452 & 0.014 & 18.8 & Mannings 1994 \\
1300 & 0.432 & 0.025 & 19.8 & Mannings 1994 \\
1300 & 0.3 & 0.0 & 3.14x3.02 & Manoj 2007 \\
1300 & 0.175 & 0.005 & 1.1x0.43 & Our data \\
2600 & 0.149 & 0.0050 & 1.42x0.88 & Our data \\
6917 & 0.029 & 0.0010 & 1.95x1.64 & Our data \\
13350 & 0.0260& 2.0$x10^{-4}$ & 3.92x3.06 & Our data \\
36000 & 0.01258 & 2.0$x10^{-4}$ & 7.2 & Skinner 1993 \\
60000 & 0.00917 & 2.0$x10^{-4}$ & 14.4 & Skinner 1993 \\
\noalign{\smallskip}\hline\hline
\end{tabular*}
\label{mwc297}
\end{table*}

\begin{table*}[h]
\caption{Extinction details for MWC 297.} 
\begin{tabular*}{1.0\textwidth}
{@{\extracolsep{\fill}}lllll}
\hline\hline\noalign{\smallskip}
Wavelength & Extinction & Extinction from envelope model & Observed flux & Dereddened flux \\
($\mu m$) & (mag) & (mag) & (mJy) & (mJy) \\
\hline\noalign{\smallskip}
0.44 & 9.24 & - & 12.2 & 60645.3 \\
0.55 & 6.97 & - & 45.8 & 28036.0 \\
0.70 & 5.23 & - & 301.0 & 37286.4 \\
0.791 & 4.26 & - & 608.9 & 30758.9 \\
0.90 & 3.34 & - & 1041.4 & 22567.3 \\
1.228 & 2.02 & - & 7726.1 & 49860.6 \\
1.235 & 2.01 & - & 5645.3 & 35818.6 \\
1.62 & 1.30 & - & 6420.2 & 21181.2 \\
1.662 & 1.24 & - & 18009.4 & 56620.4 \\
2.159 & 0.82 & - & 40507.0 & 85900.8 \\
2.2 & 0.79 & - & 36926.7 & 76571.6 \\
\noalign{\smallskip}\hline\hline
\end{tabular*}
\label{mwc297_ext}
\end{table*}

\clearpage
\newpage

\begin{table*}[h]
\caption{Table of fluxes for MWC 1080.} 
\begin{tabular*}{1.0\textwidth}
{@{\extracolsep{\fill}}lllll}
\hline\hline\noalign{\smallskip}
Wavelength ($\mu m$) & Flux (Jy) & Flux error (Jy) & Beam size ($''$) & Reference / Flux type \\
\hline\noalign{\smallskip}
0.440 & 0.022 & 0.001 & 0.8 & TYCHO-2 BAND$\_$B \\
0.550 & 0.080 & 0.003 & $<$1 & Eisner 2004 BAND$\_$V$\_$JOHNSON$\_$MORGAN \\
0.550 & 0.069 & 0.001 & 0.8 & TYCHO-2 BAND$\_$V \\
0.700 & 0.120 & 0.024 & 0.9 & USNO-B BAND$\_$R$\_$JOHNSON$\_$MORGAN \\
0.900 & 0.210 & 0.040 & 0.9 & USNO-B BAND$\_$I$\_$JOHNSON$\_$MORGAN \\
1.235 & 1.650 & 0.030 & 2 & BAND$\_$J$\_$2MASS \\
1.620 & 1.921 & 0.001 & 1 & Morel 1978 BAND$\_$H$\_$JOHNSON$\_$MORGAN \\
1.662 & 4 & 0 & 2 & BAND$\_$H$\_$2MASS \\
2.159 & 8 & 0 & 2 & BAND$\_$Ks$\_$2MASS \\
2.200 & 9 & 0 & 1 & Morel 1978 BAND$\_$K$\_$JOHNSON$\_$MORGAN \\
3.400 & 17 & 0 & 1 & Morel 1978 BAND$\_$L$\_$JOHNSON$\_$MORGAN \\
4.350 & 16.66 & 0.01 & 18.3 & BAND$\_$B2$\_$MSX6C \\
8.280 & 16.99 & 0.01 & 18.3 & BAND$\_$A$\_$MSX6C \\
12.130 & 17.91 & 0.01 & 18.3 & BAND$\_$C$\_$MSX6C \\
14.650 & 13.88 & 0.01 & 18.3 & BAND$\_$D$\_$MSX6C \\
21.340 & 16.12 & 0.01 & 18.3 & BAND$\_$E$\_$MSX6C \\
60.000 & 150.0 & 1.0 & 60 & Hillenbrand 1992 (IRAS) \\
100 & 246.2 & 0.1 & 120 & Hillenbrand 1992 (IRAS) \\
350 & $<$24.55 & - & 18.5 & Mannings 1994 \\
450 & 5 & 0 & 18.3 & Mannings 1994 \\
600 & 1.61 & 0.22 & 17.5 & Mannings 1994 \\
750 & 0.774 & 0.06 & 18.0 & Mannings 1994 \\
800 & 0.638 & 0.014 & 17.0 & Mannings 1994 \\
850 & 0.660 & 0.080 & 18.0 & Mannings 1994 \\
1100 & 0.250 & 0.030 & 18.8 & Mannings 1994 \\
1300 & 0.237 & 0.03 & 19.8 & Mannings 1994 \\
1400 & 0.0031 & 2.0$x10^{-4}$ & 1.28x0.79 & Our data \\
2700 & $<$0.0017 & - & 4.89x2.75 & Our data \\
36000 & 1.4$x10^{-4}$ & 2.0$x10^{-5}$ & 1.0 & Skinner 1993 \\
\noalign{\smallskip}\hline\hline
\end{tabular*}
\label{mwc1080}
\end{table*}

\begin{table*}[h]
\caption{Extinction details for MWC 1080.} 
\begin{tabular*}{1.0\textwidth}
{@{\extracolsep{\fill}}lllll}
\hline\hline\noalign{\smallskip}
Wavelength & Extinction & Extinction from envelope model & Observed flux & Dereddened flux \\
($\mu m$) & (mag) & (mag) & (mJy) & (mJy) \\
\hline\noalign{\smallskip}
0.44 & 6.16 & 4.24 & 21.9 & 6363.0 \\
0.55 & 4.64 & 3.19 & 79.6 & 5738.3 \\
0.55 & 4.64 & 3.19 & 68.7 & 4952.7 \\
0.70 & 3.49 & 2.40 & 119.8 & 2977.8 \\
0.90 & 2.23 & 1.53 & 209.7 & 1630.0 \\
1.235 & 1.34 & 0.92 & 1652.3 & 5662.8 \\
1.62 & 0.86 & 0.59 & 1921.1 & 4257.5 \\
1.662 & 0.83 & 0.57 & 4152.4 & 8911.5 \\
2.159 & 0.54 & 0.37 & 7825.8 & 12917.3 \\
2.20 & 0.53 & 0.36 & 8696.4 & 14141.4 \\
\noalign{\smallskip}\hline\hline
\end{tabular*}
\label{mwc1080_ext}
\end{table*}

\begin{table*}[h]
\caption{Compilation of disks masses$^1$.
} 
\nocite{ack04, pie03, man00, nat97, nat01, hen98b, hen94, dif97, pie06, man97, syl01, fue03, fue06, alo08}
\begin{tabular*}{1.0\textwidth}
{@{\extracolsep{\fill}}llllllllll}
\hline\hline\noalign{\smallskip}
Name & Spectral & Luminosity & Teff & Distance & Star mass & Age & $\lambda$ & Flux & Disk mass \\ 
 & type & (L$_\odot$) & (K) & (pc) & (M$_\odot$) & (Myr) & (mm) & (mJy) & (M$_\odot$) \\ 
\hline\noalign{\smallskip}
AB Aur & A0 & 49 & 9500 & 144 & 2.57 & 3.87 & 1.4 & 85 & 0.0278 \\ 
HD 100546 & B9 & 45 & 10500 & 103 & 2.53 & 5.59 & 1.3 & 465 & 0.0540 \\ 
HD 142527 & F6 & 69 & 6300 & 200 & 3.54 & 1.11 & 1.3 & 1190 & 1.4860 \\ 
HD 179218 / MWC 614 & B9 & 500 & 10000 & 240 & 5.02 & 0.473 & 1.3 & 71 & 0.0447 \\ 
HD 100453 & A9 & 8 & 7400 & 114 & 1.66 & 14.1 & 1.2 & 270 & 0.0709 \\ 
CD-36 10010B & F8 & 8 & 6200 & 84 & 1.63 & 9.49 & 1.3 & 142 & 0.0345 \\ 
HD 139614 & A7 & 11 & 7800 & 157 & 1.83 & 10.7 & 1.3 & 242 & 0.1293 \\ 
HD 169142 & B9 & 44 & 10500 & 145 & 2.5 & 5.69 & 1.3 & 169 & 0.0389 \\ 
HD 142666 & A8 & 8 & 7500 & 116 & 1.65 & 14.9 & 1.2 & 79 & 0.0201 \\ 
HD 144432 & A9 & 32 & 7300 & 200 & 2.26 & 3.87 & 1.3 & 37 & 0.0369 \\ 
HD 150193 / MWC 863 & A1 & 30 & 9500 & 150 & 2.25 & 6.10 & 1.3 & 45 & 0.0143 \\ 
HD 163296 & A1 & 36 & 9500 & 122 & 2.49 & 5.79 & 1.3 & 705 & 0.1485 \\ 
HD 34282 & A0 & 36 & 9500 & 400 & 2.49 & 5.79 & 1.3 & 110 & 0.2259 \\ 
HD 35187 & A2 & 34 & 9100 & 150 & 2.43 & 4.65 & 1.2 & 20 & 0.0055 \\ 
UX Ori & A3 & 42 & 8600 & 430 & 2.43 & 4.22 & 1.2 & 19.8 & 0.0481 \\ 
CQ Tau & F2 & 5 & 6800 & 100 & 1.49 & 769 & 1.3 & 143 & 0.0415 \\ 
MWC 758 & A3 & 21 & 8500 & 200 & 2.17 & 6.26 & 1.3 & 72 & 0.0471 \\ 
LkH$\alpha$259 & A9 & 107 & 7320 & 850 & 3.6 & 1.17 & 2.6 & $<$6 & $<$0.722 \\ 
WW Vul & A3 & 43 & 8600 & 550 & 2.44 & 4.19 & 1.2 & 9.1 & 0.0362 \\ 
BF Ori & A5 & 34 & 8300 & 450 & 2.34 & 4.37 & 1.3 & 6 & 0.0228 \\ 
SV Cep & A0 & 14 & 8600 & 715 & 1.91 & 11.7 & 1.3 & 7 & 0.0459 \\ 
V376 Cas & B5 & 190 & 13800 & 600 & 3.77 & 2.66 & 2.7 & $<$5 & $<$0.089 \\ 
LkH$\alpha$198 & B7 & 210 & 12000 & 600 & 3.51 & 1.49 & 2.7 & $<$5 & $<$0.1141 \\ 
VY Mon & B8 & 1400 & 10700 & 800 & 6.9 & 0.243 & 1.3 & 120 & 0.6874 \\ 
HD 97048 & A0 & 44 & 10000 & 150 & 2.53 & 4.61 & 1.3 & 452 & 0.1305 \\ 
LkH$\alpha$234 & B5 & 800 & 14500 & 1250 & 5.01 & 0.524 & 1.3 & $<$20 & $<$0.181 \\ 
T Ori & A3 & 83 & 8500 & 460 & 2.92 & 2.49 & 1.3 & $<$6 & $<$0.021 \\ 
V380 Ori & B9 & 100 & 9500 & 460 & 3.05 & 2.00 & 1.3 & $<$24 & $<$0.056 \\ 
LkH$\alpha$218 & B9 & 107 & 10700 & 1150 & 2.97 & 2.77 & 1.3 & $<$27 & $<$0.3906 \\ 
Elias 3-1 & A6 & 21 & 8000 & 140 & 2.06 & 6.11 & 2.7 & 45 & 0.1360 \\ 
BD+404124 & B2 & 5900 & 22000 & 1000 & - & - & 2.7 & $<$5 & $<$0.133 \\ 
V1686 Cyg & B5 & 1820 & 15500 & 1000 & 6.05 & 0.312 & 2.7 & $<$5 & $<$0.246 \\ 
BD+61154 & B8 & 225 & 11800 & 650 & 3.88 & 1.43 & 2.7 & 11.2 & 0.3509 \\ 
RR Tau & A4 & 150 & 8800 & 800 & 3.57 & 1.34 & 1.3 & $<$20 & $<$0.224 \\ 
HD 200775 & B2 & 1900 & 18500 & 600 & 6.9 & 0.337 & 1.3 & $<$6 & $<$0.007 \\ 
MWC 480 & A3 & 25 & 8500 & 130 & 2.19 & 5.98 & 1.4 & 235 & 0.0797 \\ 
HD 34700 & G0 & 20 & 6000 & 90 & 2.36 & 3.39 & 1.35 & 11.7 & 0.0037 \\ 
MWC 297 & B1.5 & 10700 & 24000 & 250 & 9 & 1 & 1.3 & 168.6 & 0.0323 \\ 
Z CMa & B8 & 9200 & 20000 & 930 & 12 & 0.05 & 1.4 & 24.1 & 0.2307 \\ 
R Mon & B0 & 5600 & 25000 & 800 & 8 & 1 & 1.4 & 8.3 & 0.0093 \\ 
MWC 1080 & B0 & 5600 & 25000 & 1000 & 10 & 1 & 1.4 & 2.1 & 0.0036 \\ 
MWC 137 & B0 & 15000 & 26000 & 1300 & 14 & 1 & 1.4 & 7.1 & $<$0.007 \\ 
VV Ser & A0 & 50 & 10000 & 260 & 2.68 & 3.87 & 1.3 & 1.44 & 0.0012 \\ 
LkH$\alpha$~215 & B7.5 & 5300 & 22000 & 800 & 7 & 0.4 & 1.4 & $<$1.5 & $<$0.009 \\ 
\noalign{\smallskip}\hline\hline
\end{tabular*}
\label{disksTable}
$^1$ The flux shown is the excess at 1.3 mm subtracting the free-free contribution in our sources. In MWC 297 we have used the flux of \cite{man07}. References for the mm fluxes are:
Acke et al. 2004,
Alonso-Albi et al. 2008,
Di Francesco et al. 1997,
Fuente et al. 2003, Fuente et al. 2006, 
Henning et al. 1994, Henning et al. 1998,
Mannings et al. 1997,
Mannings et al. 2000,
Natta et al. 1997,
Natta et al. 2001, 
Pietu et al. 2003,
Pietu et al. 2006,
Sylvester et al. 2001.

\end{table*}

\end{appendix}

\end{document}